\documentclass[apj]{emulateapj}

\shorttitle{Sombrero Galaxy Globular Cluster System}
\shortauthors{Spitler et al.}

\begin{document}

\title{HST/ACS Wide-Field Photometry of the Sombrero Galaxy\\Globular Cluster System\altaffilmark{1}}

\author{Lee R. Spitler}
\affil{UC Observatories / Lick Observatory, University of California, Santa Cruz, CA 95064, USA}
\affil{Centre for Astrophysics \& Supercomputing, Swinburne University, Hawthorn, VIC 3122, Australia}
\email{lspitler@astro.swin.edu.au}
\author{S\o ren S. Larsen}
\affil{European Southern Observatory, Karl-Schwarzschild-Str. 2, D-85748, Garching bei Munchen, Germany}
\affil{Astronomical Institute, University of Utrecht, Princetonplein 5, 3584 CC Utrecht, The Netherlands}
\email{s.larsen@phys.uu.nl}
\author{Jay Strader}
\affil{UC Observatories / Lick Observatory, University of California, Santa Cruz, CA 95064, USA}
\email{strader@ucolick.org}
\author{Jean P. Brodie}
\affil{UC Observatories / Lick Observatory, University of California, Santa Cruz, CA 95064, USA}
\email{brodie@ucolick.org}
\author{Duncan A. Forbes}
\email{dforbes@swin.edu.au}
\affil{Centre for Astrophysics \& Supercomputing, Swinburne University, Hawthorn, VIC 3122, Australia}
\author{Michael A. Beasley}
\affil{UC Observatories / Lick Observatory, University of California, Santa Cruz, CA 95064, USA}
\affil{Instituto de Astrof\'{i}sica de Canarias, Via Lactea, E-38200 La Laguna, Tenerife, Spain}
\email{beasley@iac.es}

\altaffiltext{1}{Based on observations made with the NASA/ESA Hubble Space Telescope, obtained [from the Data Archive] at the Space Telescope Science Institute, which is operated by the Association of Universities for Research in Astronomy, Inc., under NASA contract NAS 5-26555. These observations are associated with program \#9714.}

\begin{abstract}
A detailed imaging analysis of the globular cluster (GC) system of the Sombrero galaxy (NGC 4594) has been accomplished using a six-image mosaic from the Hubble Space Telescope Advanced Camera for Surveys.  The quality of the data is such that contamination by foreground stars and background galaxies is negligible for all but the faintest 5\% of the GC luminosity function (GCLF).  This enables the study of an effectively pure sample of 659 GCs until $\sim 2$ mags fainter than the turnover magnitude, which occurs at M$^{TOM}_{V} = -7.60\pm0.06$ for an assumed m$-$M$=29.77$.  Two GC metallicity subpopulations are easily distinguishable, with the metal-poor subpopulation exhibiting a smaller intrinsic dispersion in color compared to the metal-rich subpopulation.  

Three new discoveries include:  (1) A metal-poor GC color-magnitude trend.  (2) Confirmation that the metal-rich GCs are $\sim17\%$ smaller than the metal-poor ones for small projected galactocentric radii (less than $\sim2$ arcmin).  However, the median half-light radii of the two subpopulations become identical at $\sim$ 3 arcmin from the center.  This is most easily explained if the size difference is the result of projection effects.  (3) The brightest (M$_V < -9.0$) members of the GC system show a size-magnitude upturn where the average GC size increases with increasing luminosity.  Evidence is presented that supports an intrinsic origin for this feature rather than a being result from accreted dwarf elliptical nuclei.  In addition, the metal-rich GCs show a shallower positive size-magnitude trend, similar to what is found in previous studies of young star clusters.
\end{abstract}

\keywords{galaxies: spiral - galaxies: star clusters - galaxies: individual (\objectname{M104, NGC 4594})}

\section{Introduction}
Extragalactic globular cluster (GC) systems provide an observational probe of galaxy formation and evolution.  GCs may have formed even before galaxies were fully assembled, as well as during subsequent star formation episodes \citep[see references in ][]{bs06}.  Since the Milky Way serves as a standard against which other GC systems are compared, it is particularly important to study GC systems in other spiral galaxies to establish whether the Milky Way GC system is typical.   

The study of spiral GC systems is complicated by their generally smaller GC populations and uncertainties in the internal extinction conditions. Nonetheless, data from the Hubble Space Telescope's (HST) Wide-Field Planetary Camera 2 (WFPC2) have helped establish the properties of GC systems in several spiral galaxies, including detailed studies of individual galaxies \citep{kea99,lfb01,bh01,l00} as well as limited-scale surveys \citep{fbl01,l02,gea03,cwl04}.  While these efforts illustrated the important role of space-based telescopes in extragalactic GC studies, large GC samples have been difficult to obtain for all but the most densely populated systems due to the small field of view of the WFPC2.  In contrast, the HST Advanced Camera for Surveys (ACS) has higher sensitivity and doubles the areal coverage. It is thus a powerful tool for global studies of GC systems, as demonstrated by a number of ACS surveys of such systems in early-type galaxies \citep{pgea05,sea05B,hea05}.

\begin{figure*}
\plotone{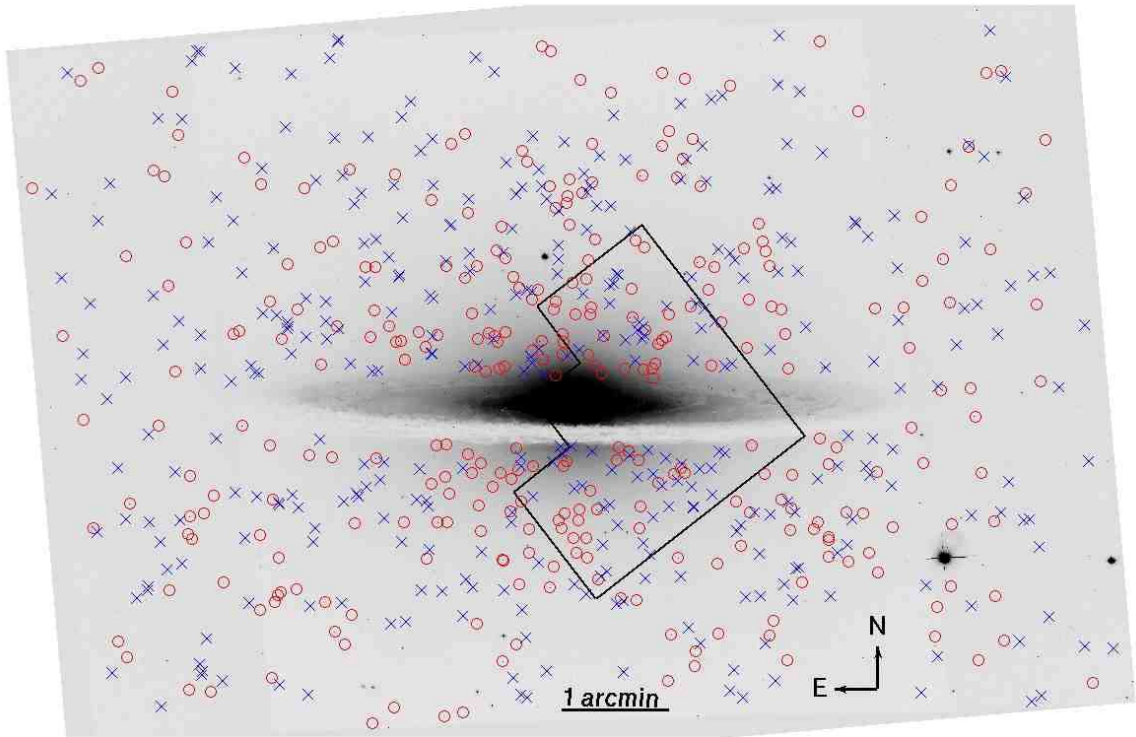}
\caption{Sombrero galaxy image mosaic from the Hubble Space Telescope Advanced Camera for Surveys.  The cross and circle symbols illustrate the locations of globular clusters belonging to metal-poor and metal-rich subpopulations, respectively.  Those objects found within the projected dust lanes were not included in the analysis to avoid uncertain extinction conditions and confusion with HII regions.  The region covered by a HST WFPC2 pointing used in an earlier Sombrero GC system study by \citet{lfb01} is shown.\label{figsomb}}
\end{figure*}

Figure~\ref{figsomb} shows NGC 4594, the ``Sombrero galaxy'', which is a nearby ($9.0\pm0.1$~Mpc or m$-$M$=29.77\pm0.03$, see $\S\ref{lfsection}$ for derivation), nearly edge-on (i=84\degr), spiral galaxy (type Sa) with a populous GC system.  The Sombrero is the most massive member of a small galaxy group.  While its classification as a spiral galaxy is questionable \citep[e.g. ][ call it an S0]{rz04}, the Sombrero does appear to be an intermediate case between elliptical and late-type spiral galaxies, therefore a unique system to analyze.

The GC systems of both spiral and elliptical galaxies generally possess at least two GC subpopulations characterized as metal-poor (blue) and metal-rich (red) GCs \citep{bs06}.  Due to the notorious metallicity-age degeneracy \citep{w94}, ages and metallicities cannot be independently constrained by broad-band colors alone.  Ages derived from Keck spectra of 14 Sombrero GCs, suggest that both subpopulations are $\sim 11-12$ Gyr old \citep{lea02}, as is the case in the Milky Way GC system.  \citet{hea02} used VLT spectra to derive mean metallicities of [Fe/H]$_{blue}=-1.7$ and [Fe/H]$_{red}=-0.7$ for the Sombrero's two GC subpopulations.  A ground-based, wide-field imaging study (R$_{gc} < $ 20\arcmin) of the Sombrero by \citet{mea02} showed that the red GCs are distributed spherically and suggested their association with the galaxy's bulge rather than its disk.  \citet{rz04} used a $36\arcmin\times36\arcmin$ wide-field, ground-based image to show that the GC surface density drops to zero at a projected radius of $\sim19\arcmin$ ($\sim50$~kpc).  They estimated a total GC population of 1900 and calculated a specific frequency, S$_N=2.1\pm0.3$, where S$_N = $N $\times10^{0.4(M_V + 15)}$ \citep{hv81}.  \citet{lfb01} found that Sombrero's red GCs are, on average, $\sim$30\% smaller in size than its blue GCs. 

\section{Data Reduction}\label{datareduction}

The Hubble Heritage Project sponsored a six-pointing, ACS mosaic of the Sombrero Galaxy (PI: Keith Noll, HST PID 9714).  Each pointing consisted of $4\times$675s exposures with the F435W (B) filter, $4\times$500s with the F555W (V) and $4\times$350s with the F625W (R).  Figure~\ref{figsomb} illustrates the $\sim600\arcsec\times400\arcsec $  region covered by the image mosaic.  Each pointing was analysed separately.  Raw ACS images were prepared for analysis with the standard STScI ``On-the-Fly'' data reprocessing system.  This data pipeline made use of the latest calibration files (e.g. dark current reference file) to prepare raw data and corrected the significant geometric distortion in the ACS by ``drizzling'' together multiple exposures.  The calibrated, drizzled data products were downloaded and prepared for object detection by removing diffuse galaxy light via the subtraction of a median filtered image.  To detect GC candidates on the background-subtracted images, the FIND task in the DAOPHOTII aperture photometry software application \citep{ste92} was used.  An intentionally low detection threshold ($3\sigma$) was used to maximize the chances of finding a faint object.  Spurious detections were effectively removed by matching up detected objects across the three bands.

Aperture photometry was carried out on the original images using the PHOT/DAOPHOTII routine with a 5 pixel radius and individual sky values calculated from a 10 pixel wide annulus that began 30 pixels from the object center.  A larger aperture size was not used because it would have contributed more noise from the sky than signal from the object.  To compensate for the light found just outside the 5-pixel aperture, the median aperture difference (5$-$10 pixels) from candidate GCs was subtracted from all objects.  Additional aperture corrections were applied according to the published 10 pixel ($0.5\arcsec$) to nominal infinite values from the ACS photometric performance and calibration \citep{siea05}.  Although ignored in the present study, a systematic error is expected on the 10 pixel to infinity corrections because, unlike the synthetic stars used to calculate these corrections, extended sources will show light beyond the 10 pixel aperture radius.  From a few extremely isolated GCs in a B-band image, this effect is estimated to be slight ($\sim0.03$ mag).  

Galactic foreground extinction corrections were applied using the reddening values, E(\bv), which corresponded to the center of each of the six pointings in the DIRBE dust maps of \citet{sfd98}.  To convert the E(\bv) values to the corresponding extinction values (A$_M$) the reddening conversion factors from Table~14 of \citet{siea05} were used assuming a G2 spectral energy distribution.  Table~\ref{tbl-corr} summarizes the corrections applied for each filter.  Corrected photometry was then transformed from the ACS system to the Johnson BV and Cousins R systems using the coefficients and technique described in \citet{siea05}.

Globular cluster sizes were calculated using the ISHAPE algorithm \citep{l99}.  This algorithm estimates the half-light radius (R$_{hl}$) by convolving a point spread function (PSF) with an analytic profile, varying the FWHM until the best match to the object profile is found.  At the distance of the Sombrero, typical GCs (R$_{hl}=2.5$pc) will show angular radii of $\sim 0.06\arcsec$, which is slightly larger than the pixel scale of $0.05\arcsec$ pixel$^{-1}$.  After testing three different PSF derivation methods (described in the following paragraph) an empirical PSF was adopted for estimating object sizes.  The final analysis made use of mean sizes derived from the three available bands.  See $\S\ref{sizeselect}$ for details of combining the size information from the three bands, size selection criteria, and the estimated size accuracy.

\begin{figure}
\center
\includegraphics[scale=0.4,angle=0]{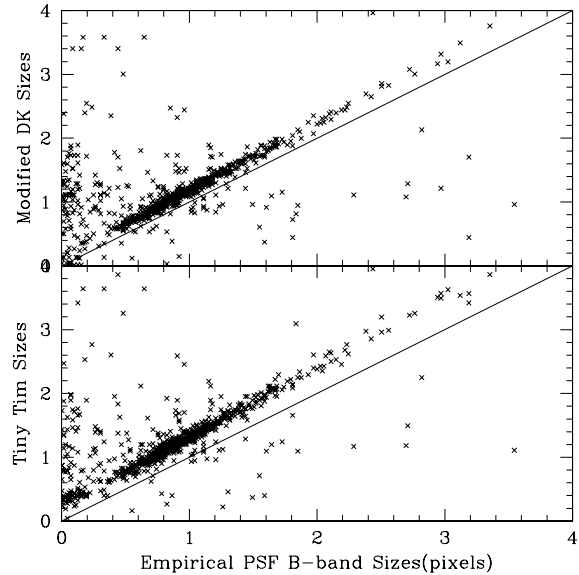}
\caption{Comparison of ISHAPE F435W size measurements using three different PSFs.  In both panels, the sizes calculated with an empirical PSF are on the abscissa.  These values are compared to the same objects whose sizes were estimated with Tiny Tim PSFs convolved with a modified DK and the original Tiny Tim PSFs on the top and lower panels, respectively.  Lines are 1:1 relations.  Empirical PSFs were adopted for the size analysis because foreground stars show the expected zero sizes.\label{figcmpr2size}}
\endcenter
\end{figure}

Model PSFs for the HST cameras are generated with the Tiny Tim program, which also supplies a charge diffusion kernel (DK) that must be convolved with the PSF after it is re-binned for comparison with an object profile.  Since the PSF varies across the two ACS CCDs, a $21\times12$ grid of Tiny Tim PSFs was generated for each CCD.  ISHAPE was directed to convolve these PSFs with a \citet{k62} model (concentration parameter fixed at $r_t/r_c = 30$) and estimate the FWHM of the object profiles on the F435W images.  This method yielded no objects with the zero half-light radius that would be expected for foreground stars, suggesting the object profiles are broadened compared to the Tiny Tim models, likely from the ``drizzling'' process.  

Two methods were explored to resolve this problem. The first used an empirical PSF constructed from bright isolated stars on the ACS images.  The second method involved the creation of a broader DK to effectively re-distribute the light away from the center of the Tiny Tim PSFs, thereby accounting for the drizzle broadening.  This broader DK was created by uniformly populating a blank pre-drizzle image (``flt'' ACS data suffix) with a number of isolated pixels with non-zero values.  After these images were drizzled, the broadened signal from the non-zero pixels was averaged and convolved with the original DK, producing a broader DK.  Figure~\ref{figcmpr2size} compares the estimated R$_{hl}$ using the original and modified DK methods with the empirical-PSF sizes.  Compared to the empirical values, sizes derived from the broader DK are larger by approximately 0.3 pixels. The original method produced half-light radii of objects resembling GCs that are larger by $\sim0.5$ pixels relative to the empirical sizes.  Both the modified DK and original methods produced non-zero sizes for those objects that appear to be stellar in the empirical sizes.  Therefore the empirical PSF-derived sizes were adopted since these are the least broadened and stars are readily identifiable.

\subsection{Globular Cluster Selection Criteria and Contamination} \label{selcrit}

Selection criteria based upon magnitude, color, size, shape, and location were used to construct a catalogue of Sombrero GC candidates.  A description of the limits imposed on the photometric properties of all detected objects follows in the next subsection.  In $\S\ref{sizeselect}$, the derivation of sizes is described and how they were employed in the selection process.  Section~\ref{contamin} addresses contamination in the Sombrero GC dataset.

\subsubsection{Photometry Selection} \label{photoselect}

\begin{figure}
\center
\includegraphics[scale=0.4,angle=0]{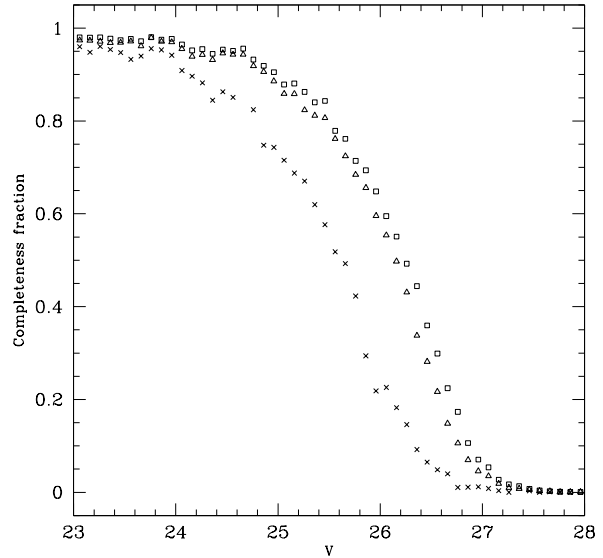}
\caption{Completeness values from artificial object detection rates.  Square, triangle, and cross symbols represent completeness values from the F555W (V) band images for the entire, outer ($1.5\arcmin \leq$ R$_{gc} < 3\arcmin$), and inner (R$_{gc} < 1.5\arcmin$) image regions, respectively.\label{figcomp}}
\endcenter
\end{figure}

Objects within the image region containing the dust lanes were excluded from the GC candidate catalogue (see Figure~\ref{figsomb}) to avoid HII regions, young massive clusters, and uncertain extinction conditions.  Photometric completeness tests were performed by generating artificial objects (using ADDSTAR/DAOPHOTII) modelled upon an empirical PSF generated from isolated GCs.  Colors of these artificial objects reflected the mean observed GC colors.  Five thousand artificial objects with the same luminosity were randomly distributed across the six pointings for each 0.1 magnitude interval between 18 and 28 mag.  The same data reduction procedure as described in $\S2.0$ (except size analysis) was carried out on these artificial fields and the resulting detection rates are shown in Figure~\ref{figcomp}.  The R$_{gc}$ of the artificial objects were recorded during the tests to obtain a more suitable completeness function when analyzing the GC luminosity function (GCLF) at different galactocentric distances.  The $50\%$ completeness level from the combined six pointings is V $\sim 26.2$.

\begin{figure}
\center
\includegraphics[scale=0.33,angle=-90]{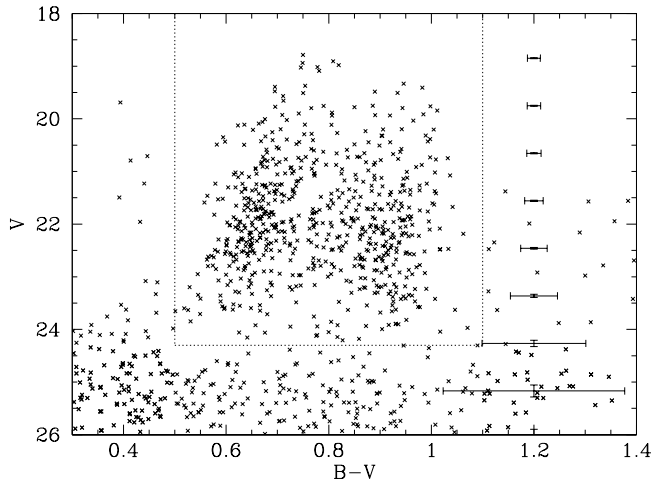}
\caption{Color-magnitude diagram illustrating the color and magnitude cuts employed.  Candidate GCs in the plot are selected on shape and size, as well as a visual inspection.  Horizontal error bars represent the average color photometric errors and vertical error bars indicate average magnitude errors for eight magnitude intervals.  Dotted vertical lines at B$-$V = 0.5 and B$-$V = 1.1 indicate the color selection limits, while the horizontal line at V $=24.3$ represents the faint magnitude limit.  The faint limit excludes the increasing amount of contamination which begins to appear at fainter magnitudes.\label{figrawcmd}}
\endcenter
\end{figure}

A faint magnitude limit of V $ = 24.3$ (at the $95\%$ completeness level) was chosen to exclude contamination that begins to contribute very significantly at fainter magnitudes, as shown in Figure~\ref{figrawcmd} and discussed in Section~\ref{contamin}.  This magnitude limit allows access to GCs $\sim 2$ mag below the observed turnover magnitude of V $= 22.17\pm0.06$, or approximately $95\%$ of the expected GC population (see $\S\ref{lfsection}$).  Since the number of objects with magnitudes near this limit is so small, essentially no selection bias is introduced by imposing a selection criteria on a single band to a sample which spans a $\sim 0.5$ range in colors (i.e. the number of candidate GCs increases by a negligible amount if B-band rather than a V-band magnitude limit is used).

\begin{figure}
\center
\includegraphics[scale=0.4,angle=0]{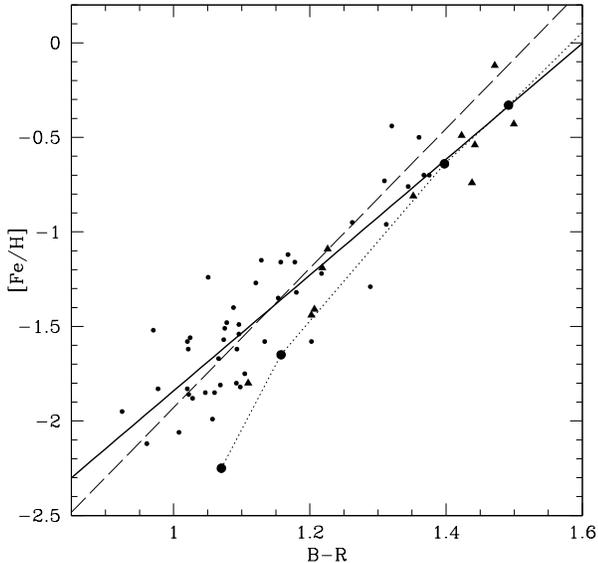}
\caption{Diagram illustrating the B$-$R color to metallicity transformation.  The transformation is represented by the solid line and was derived from a linear fit to Milky Way (MW) GCs with Galactic reddening values of E(B$-$V) $<0.3$ from the \citet{h96} catalogue (circles) and the metallicity values derived from Sombrero GC spectroscopy \citep[triangles; ][]{lea01}.  For comparison, the color-metallicity transformations from \citet[][ dashed line]{baea02} and the \citet[][ dotted line]{bc03} 13 Gyr models are presented.  The former is consistent with the MW GCs, but deviates from the Sombrero data points in the metal-rich GC region, while the later is offset from the metal-poor GCs. \label{figcolormet}}
\endcenter
\end{figure}

Objects were selected as GCs if their colors fell within the ranges $0.9 < $ B$-$R $ < 1.7$ and $0.5 < $ B$-$V $ < 1.1$.  As shown in Figure~\ref{figrawcmd}, these color limits correspond to the observed GC colors, with a small ``buffer'' to allow for some photometric scatter among the faintest GCs.  To convert broad-band colors to metallicities a straight line was fitted to the [Fe/H] and color values of Milky Way GCs with E(B$-$V) $< 0.3$ \citep{h96}.  To supplement the metal-rich end, spectroscopic metallicities derived from Sombrero GCs \citep{lea02} were paired with the ACS photometric information and included in the linear fits.  Figure~\ref{figcolormet} shows this data, which yielded the following color-metallicity transformations:  [Fe/H] $ \approxeq -4.90 + 3.06$(B$-$R), [Fe/H] $\approxeq -4.45+ 4.39$(B$-$V) and [Fe/H] $\approxeq -5.64 + 9.43$(V$-$R).  The GC color selection limits approximately correspond to the following metallicity ranges:  $-2.1 <$ [Fe/H]$_{(B-R)}$ $< +0.3$ and $-2.3 <$ [Fe/H]$_{(B-V)}$ $< +0.4$. 

Detections were only adopted as GCs if their FIND/DAOPHOTII roundness values were between $-0.45$ to $0.45$ and their sharpness values were between 0.55 and 0.85.  Roundness and sharpness thresholds were based upon the inspection of all objects that resembled GC in color and size.  Finally, GC candidates were inspected for any obvious contaminants (e.g., HII regions, disky galaxies) and for duplicate objects lying on the six-image mosaic overlap regions.

\subsubsection{Size Estimates and Selection} \label{sizeselect}
As described in $\S\ref{datareduction}$, object sizes were determined using an empirical PSF and the profile-fitting routine ISHAPE.  To reduce noise from the fitting process, three size values were determined for each object by running ISHAPE on each of the B, V and R band images.  The final size was taken to be the mean of the three size values, which first underwent an iterative sigma-clipping algorithm to remove outliers and systematics as is described in the following paragraph.  

\begin{figure}
\center
\includegraphics[scale=0.4,angle=0]{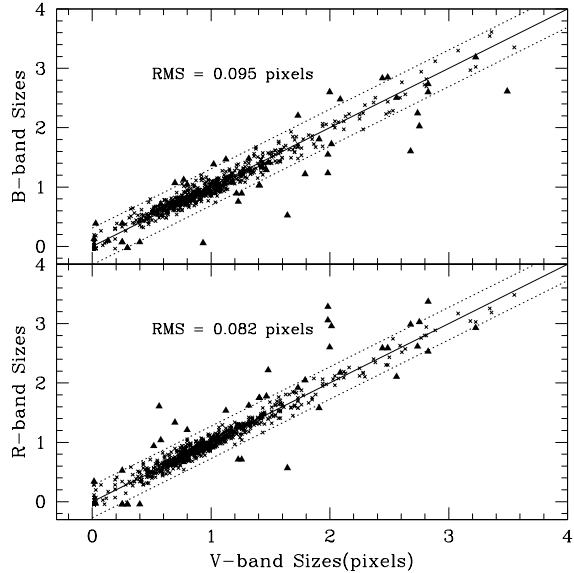}
\caption{Comparison of sizes from B-band (top) and R-band (bottom) images compared with V-band sizes.  Both have undergone an iterative sigma-clipping procedure to remove systematics and outliers, as is described in the text.  Triangle symbols represent the objects that fell outside of the $3\sigma$ regions indicated by dotted lines.  Solid line is a 1:1 line for comparison.\label{figbandsizes}}
\endcenter
\end{figure}

The V band sizes were arbitrarily chosen to be the reference system.  Using the objects which passed the aforementioned selection criteria, systematic differences between the bands were estimated by fitting a line to the V versus B and V versus R size values.  These fits were used to remove the systematics by forcing the B and R sizes to show a 1:1 trend with the V sizes.  Objects falling more than $3\sigma$ from the two 1:1 lines were discarded from further analysis.  A line was fit to the modified dataset to gain a more accurate understanding of the systematics.  After applying a correction to the B and R sizes with the improved systematic estimate, another $3\sigma$ clip was applied to each of the final 1:1 lines.  For the remaining objects the mean of the resulting size values was adopted as the final size.  

\begin{figure}
\center
\includegraphics[scale=0.4,angle=0]{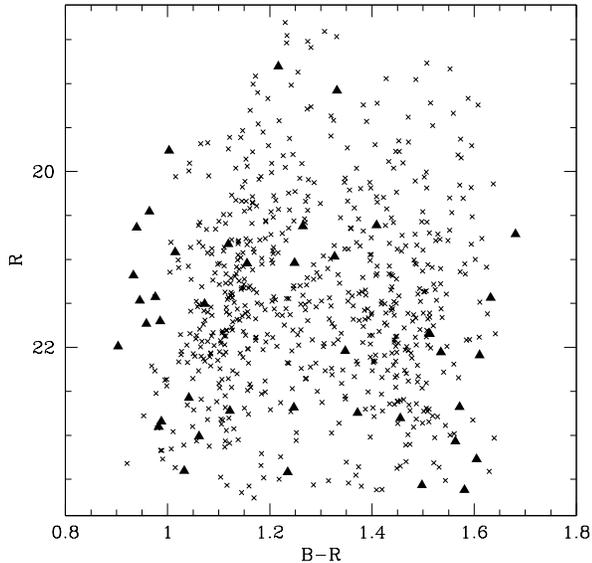}
\caption{Color-magnitude diagram with the crosses representing the final GC dataset and the triangles the objects removed by the size sigma-clipping described in the text.  The objects tend to be faint and show colors unlike typical GCS.}\label{figsizecmd}
\endcenter
\end{figure}

Figure~\ref{figbandsizes} compares the size values of the three bands, after the above procedure was applied.  The majority of the objects that were removed in the sigma clipping algorithm tend to be fainter objects, as shown in Figure \ref{figsizecmd}.  A significant difference between the observed object sizes in at least one of the bands likely indicates the object is not a typical old stellar system, an idea that is supported by the fact that a significant portion of these objects deviate from typical GC colors (Figure~\ref{figsizecmd}).

\begin{figure}
\center
\includegraphics[scale=0.4,angle=0]{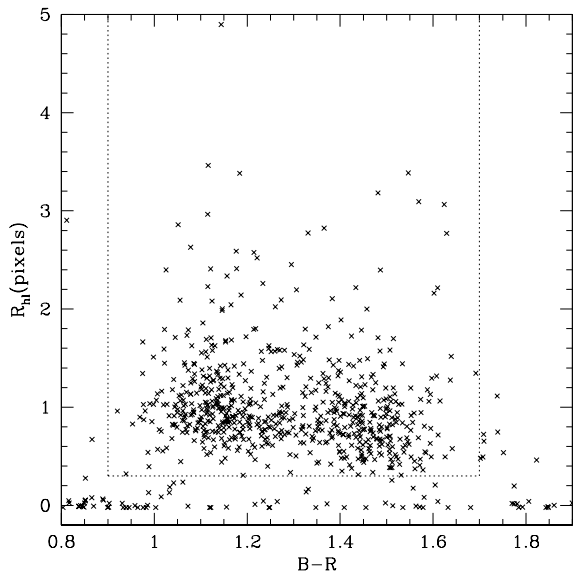}
\caption{Object size versus B$-$R illustrating the adopted size and color selection limits.  Data are selected on sharpness, roundness, luminosity, and a visual inspection.  Those objects below the horizontal dotted line at R$_{hl}=0.3$~pixels are likely foreground stars.  Dotted vertical lines represent B$-$R color limits of 0.9 and 1.7. \label{figrawsize}}
\endcenter
\end{figure}

Least-squares RMS values of the final size fits are 0.095 and 0.082 pixels for the V versus B and V versus R sizes, respectively.  Thus the sizes are accurate at approximately the 0.1~pixels (or 0.2~pc) level.  To remove stellar contaminants from the dataset, a lower-limit of 0.3~pixels was enforced to account for any sizes that deviated by $3\sigma$ from a size of zero, as shown in Figure~\ref{figrawsize}.  Objects with ISHAPE ellipticities ($\epsilon=1-b/a$) greater than 0.5 were also discarded as likely background galaxies.

The completeness tests presented at the beginning of this section did not account for the size selections, therefore these completeness functions are not perfect representations of the complete Sombrero GC selection process.  The true completeness functions likely differ by a small amount, since the objects removed in the sigma clipping algorithm tend to be unlike GCs (Figure~\ref{figsizecmd}) and the number of stars predicted and found in the mosaic make up only a small fraction of the data sample (see the following subsection).

\subsubsection{Contamination Estimates} \label{contamin}

\begin{figure}
\center
\includegraphics[scale=0.4,angle=0]{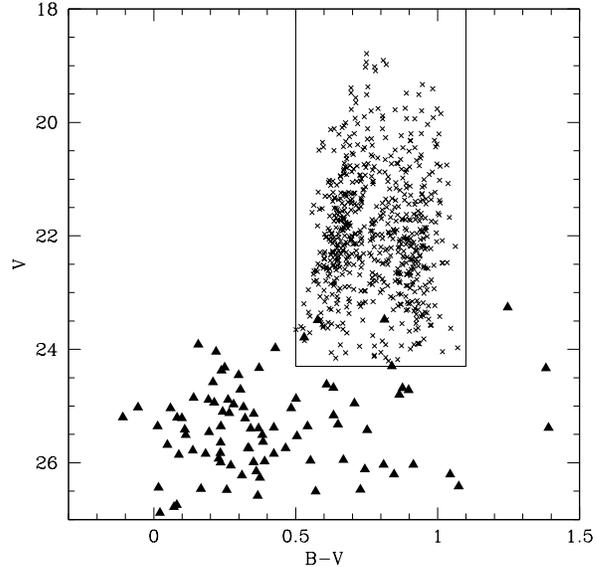}
\caption{Color-magnitude diagram with crosses representing the final Sombrero GCs.  The triangles are the objects found on a control field.\label{figcmdcont}}
\endcenter
\end{figure}

An unknown quantity of contamination from background galaxies and foreground stars is present in this dataset.  While the resolving capabilities of the ACS instrument make identifying unwanted object more straight-forward, it is nevertheless important to estimate how much contamination is missed.  Reducing an identical set of ACS images taken just beyond the ``edge'' of the Sombrero GC system is the ideal method for estimating the contribution from contamination.  Lacking such a control field, B and V images (2400~seconds~in B and 1120s~in V) from the Hubble Deep Field North of the GOODS survey \citep[HST ID 9583;][]{g04} were utilized.  A single region of the survey was reduced in exactly the same manner as the Sombrero mosaic and sizes were determined on the B image bands.  The images were examined and found to have few Sombrero GC-like objects when the selection criteria were applied, as shown in Figure~\ref{figcmdcont}.  Even when one increases the observed objects found in the control field by a factor of six to account for the six pointings in the ACS mosaic, the number of background objects classified as Sombrero GCs would be only 24 objects, or $\sim4\%$ of the final Sombrero GC sample.  This investigation also supports the faint Sombrero magnitude limit, which was motivated by an obvious increase in the background contamination at V $\sim24.3$ (see Figure~\ref{figrawcmd}), corresponding to the objects detected in the control field.

Since this control field was originally selected to avoid stars, it is necessary to estimate the contamination from foreground stars.  From the \citet{bs80} Galaxy model, approximately 40 stars with the same apparent magnitude and B$-$V colors as the Sombrero GCs are estimated to be within the ACS mosaic region.  After applying all photometric selection criteria to the objects detected in the Sombrero field, approximately 41 objects are classified as foreground stars because they fall below the lower size limit of R$_{hl}=0.3$ pixels (i.e. there are 41 objects within the color limits and below the vertical line in Figure~\ref{figrawsize}).  This number is consistent with the predicted number of stars, thus the Sombrero GC sample likely contains a negligible amount of contamination from foreground stars.

\section{Results and Discussion}

\subsection{Color-Magnitude Distribution}\label{sectioncmd}

\begin{figure*}
\center
\includegraphics[scale=0.63,angle=-90]{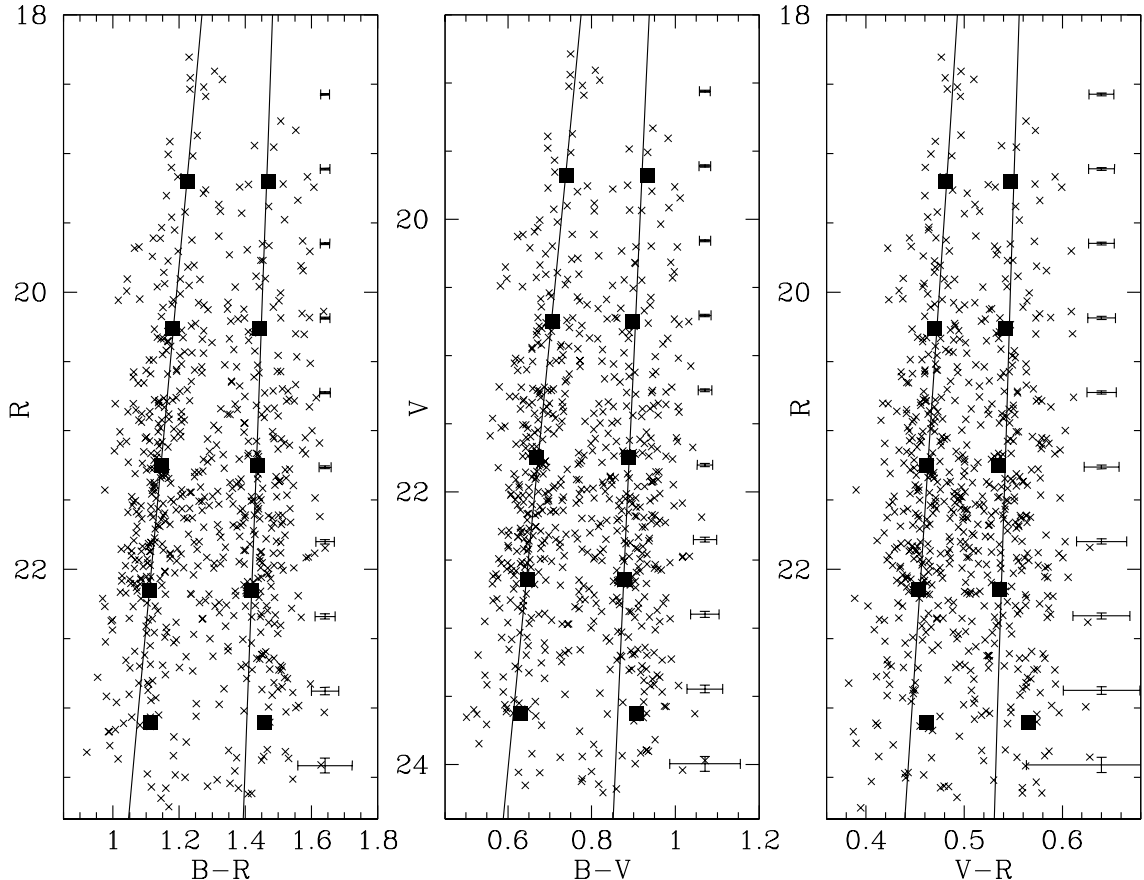}
\caption{Color-magnitude diagrams showing the globular clusters of the Sombrero galaxy.  Color/metallicity peaks, dispersion values and subpopulation proportions for each distributions are presented in Table~\ref{tabcolor}.  Error bars are the same as in Figure~\ref{figrawcmd}.  As discussed in $\S\ref{sectiontilt}$, a color-magnitude trend is observed in the blue subpopulation where brighter GCs show, on average, redder colors.  This is illustrated with solid lines from least-squares fits to peak colors of \textit{only} the four brightest magnitude bins, which are represented by squares.\label{figgoodcmd}}
\endcenter
\end{figure*}

Figure~\ref{figgoodcmd} shows the B$-$R, B$-$V and V$-$R color-magnitude diagrams of the 659 objects that met the GC selection criteria.  To parametrize the GC color distribution a Bayesian univariate normal mixture analytical method \citep{rg97} was used via a software routine named ``Nmix''\footnote{\url{http://www.stats.bris.ac.uk/$\sim$peter/Nmix/}}.  Peak colors of B$-$R = 1.15 and 1.44 ([Fe/H] $= -1.38$ and $-0.49$) were found with 53\% of the GCs assigned to the blue subpopulation.  As shown in Table~\ref{tabcolor}, the GC subpopulation peak metallicities and blue proportions are similar to the same values from the B$-$V and V$-$R color distributions.  A KMM test \citep{abz94} performed on the B$-$R colors supports a bimodal distribution over a unimodal one at the 99.9\% confidence level.  \citet{rz04} detected a Sombrero bimodal metallicity distribution with 59\% blue GCs and subpopulation peak B$-$R values consistent with these results.  Since their ground-based study extended to the ``edge'' of the GC system at R$_{gc}=19\arcmin$ (while the ACS mosaic has R$_{gc}\lesssim6\arcmin$), and because blue clusters tend to dominate the halo, their larger percentage of blue GCs is not unexpected.  In the subsequent analysis, blue GCs are taken to be those with B$-$R colors $\leq 1.30$, and red GCs are those with B$-$R $>$ 1.30.  The only exception to this rule are the two brightest GCs with B$-$R $>$ 1.30.  It is assumed these are in fact ``metal-poor GCs'' that happen to lie at the extreme of a metallicity-mass trend (presented in Section~\ref{sectiontilt}) and therefore show redder colors compared to typical metal-poor GCs.

Using a conversion from B$-$R to V$-$I from Milky Way GCs \citep{ff01}, the blue and red subpopulation peaks approximately correspond to V$-$I = 0.93 and 1.13, respectively.  \citet{lfb01} found on V and I-band WFPC2 images, color peaks of V$-$I = 0.96 and 1.21.  When compared to the GC peak V$-$I color versus galaxy luminosity relations of \citet{sbf04}, the ACS Sombrero blue GCs fall directly onto the relation, while the red GC peak is slightly bluer than expected.  The ACS red peak is likely different from the past studies because the red part of the color transformation is poorly constrained due to the small numbers of red Milky Way GCs.

To quantify the intrinsic dispersions of the subpopulations, the sample was limited to V $<22.0$ and Nmix was used to estimate the Gaussian dispersions:  $\sigma^{B-R}_{blue} = 0.08$ and $\sigma^{B-R}_{red} = 0.10$.  These $1\sigma$ color ranges, when converted to metallicity, are consistent with the dispersions derived when the color values were first converted to metallicity:  $\sigma^{[Fe/H]}_{blue} = 0.23$ and $\sigma^{[Fe/H]}_{red} = 0.28$.  Supporting this analysis, the same procedure applied to the B$-$V and V$-$R color distributions yielded similar metallicity dispersions as shown in Table~\ref{tabcolor}.

Other studies have similarly exploited the low photometric uncertainty of the ACS data and estimated the intrinsic subpopulation dispersions \citep{sea05B,hea05,pgea05}.  As found here, each of these studies detected a larger color spread in the metal-rich subpopulation compared to the metal-poor GCs.  However, lacking a standardized color-to-metallicity transformation needed for a physical interpretation, it is difficult to make conclusive statements about the intrinsic dispersions of the GC subpopulations.  For instance, the present analysis agrees with \citet{hea05} in that a \textit{smaller} metallicity dispersion is inferred in the metal-poor subpopulation, but a non-linear color-to-metallicity transformation in \citet{pgea05} yielded a \textit{larger} metallicity dispersion among the metal-poor GCs compared to the metal-rich GCs.  \citet{pgea05} briefly explore the physical implications of the subpopulation metallicity dispersions and similarly express their reservations about making too many physical interpretations until a definitive color-to-metallicity transformation is established.

\subsection{Luminosity Function} 
\label{lfsection}

\begin{figure}
\center
\includegraphics[scale=0.4,angle=0]{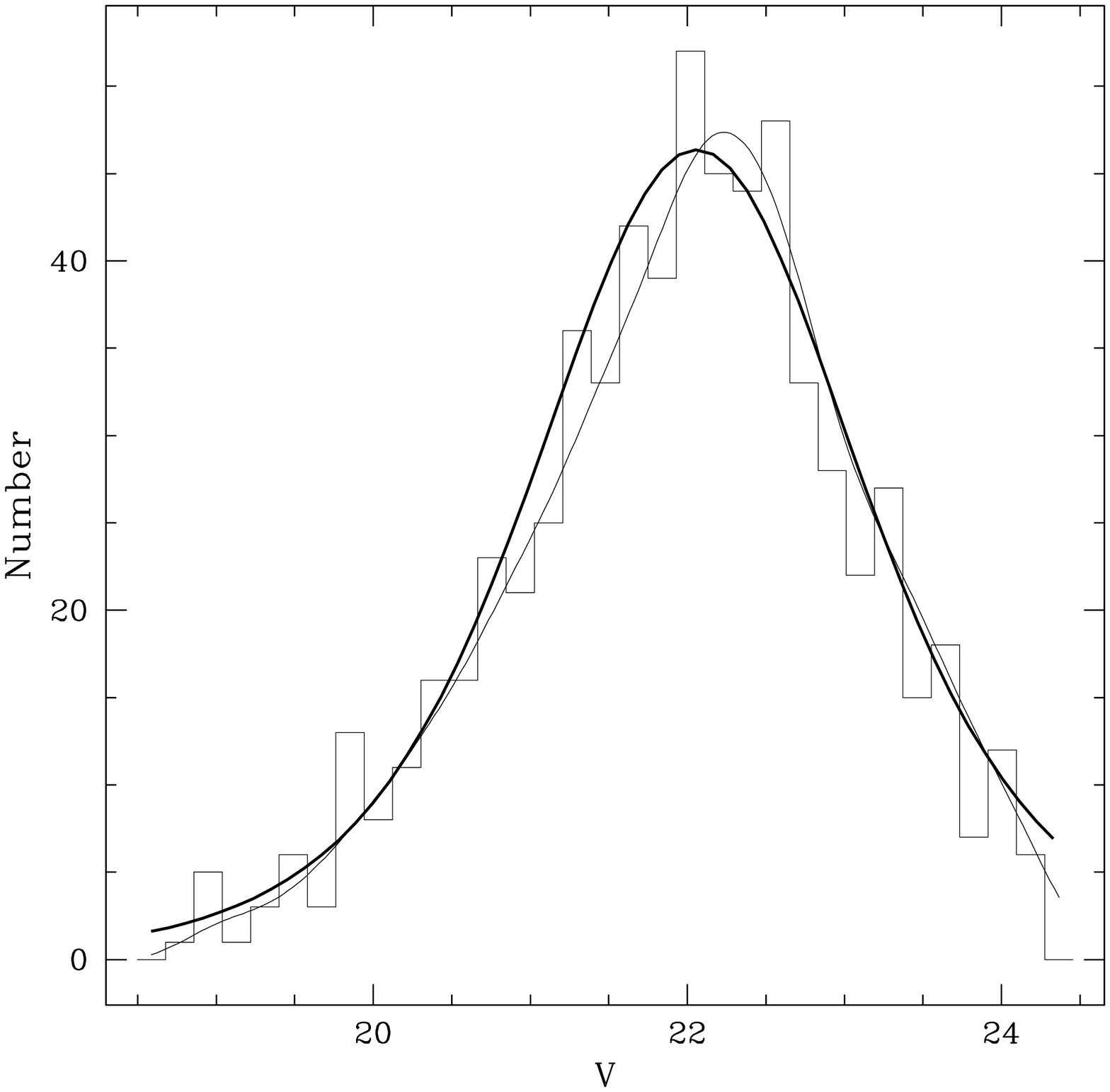}
\caption{Observed V-band GC luminosity function, with the best-fitting Student-$t5$ distribution over-plotted as a thick curve (turnover magnitude: m$_V^{TOM}= 22.06\pm0.06$; dispersion: $\sigma_{t5} = 1.08\pm0.05$).  The thin curve represents the smoothed density estimate of the distribution, the peak of which is taken to be the Sombrero TOM:  m$_V^{TOM}= 22.17$ or M$_V^{TOM}=-7.60$.  From artificial field tests it is estimated the 95\% completeness level is V $\sim24.3$, thus the ACS mosaic enables the detection of nearly all GCs brighter than the faint magnitude limit of V $=24.3$.\label{figlfv}}
\endcenter
\end{figure}

Figure~\ref{figlfv} shows the Sombrero V-band GCLF overlaid with a Student-t5 ($t_5$) distribution fit from the maximum-likelihood algorithm of \citet{sec92}.  This algorithm accounts for completeness and photometric errors.  The long exposure time and the resolving power of the ACS camera enabled contaminants to be identified and eliminated with unprecedented accuracy, resulting in an essentially pure sample of Sombrero GCs.  By integrating the best-fitting Sombrero GCLF $t_5$ distribution, it is estimated that only 5\% of the GCs are missed from the magnitude limit of V $< 24.3$.  Such quality and the large sample allows the faint end of the GCLF to be directly examined, showing for the first time that the \textit{entire} Sombrero GCLF follows a $t5$ distribution.

The location where the GCLF distribution peaks, the turnover magnitude (TOM), is a useful quantity to compare the GCLFs of different galaxies.  The apparent $V$-band TOM from a $t_5$ fit, m$^{TOM}_{V} = 22.06\pm0.06$, is the equivalent to a Gaussian TOM estimate:  m$^{TOM}_{V} = 22.03\pm0.06$.  A $t_5$ distribution is adopted because its shape appears to better represent the wings of the GCLF (confirmed with a KS test).  Comparing the $t_5$ fit to the GCLF histogram in Figure~\ref{figlfv}, it is apparent that the fit may not accurately represent the TOM.  Indeed, the peak of a smoothed density kernel estimate shown in the same Figure suggests a slightly fainter TOM might be more appropriate.  The peak in density estimate corresponds to an apparent magnitude of m$^{TOM}_{V} = 22.17$, which is taken to be the Sombrero GCLF TOM.

TOM values derived from the two other available bands and color conversions from the Sombrero mean GC colors are consistent with the TOM values found in the V band.  The ACS $t_5$ TOM is consistent with a previous WFPC2 Sombrero turnover estimate of m$^{TOM}_{V} = 22.10\pm0.15$ \citep{lfb01}, but the $t_5$ dispersion is broader than the WFPC2 estimate:  $\sigma^{ACS}_{t5} = 1.08\pm0.06$, $\sigma^{WFPC2}_{t5} = 0.81\pm0.10$.  The difference between the past result is likely a result of small number statistics and/or an increased amount of contamination at fainter magnitudes in the WFPC2 dataset.  This is demonstrated by a ``jump'' in the estimated WFPC2 GCLF dispersion to $\sigma^{WFPC2}_{t5} = 1.02\pm0.12$ when the magnitude limit is increased by half of a magnitude \citep{lfb01}.

\begin{figure}
\center
\includegraphics[scale=0.4,angle=0]{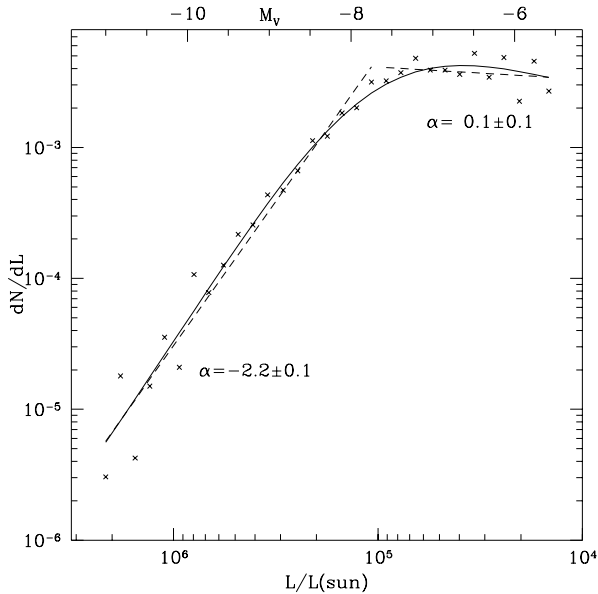}
\caption{GC luminosity function in units of solar luminosity. A Student-t5 distribution fit from the \citet{sec92} algorithm is represented as a solid line.  Dashed lines are power-law fits ($dN(L_{V})/dL_{V} = \beta L_{V}^\alpha$) to two regions of the distribution, with the corresponding power-law exponents indicated next to each of the lines.\label{fignormlf}}
\endcenter
\end{figure}

Figure~\ref{fignormlf} shows a normalized form of the Sombrero GCLF in units of solar luminosity.  The TOM corresponds to the bend in the distribution.  A power-law fit of the form $dN(L_{V})/dL_{V} = \beta L_{V}^\alpha$ yields an exponent of $\alpha=-2.2\pm0.1$ for the points brighter than the bend.  This exponent is consistent with observations of young star clusters luminosity functions in spiral galaxies \citep{l02}.  Power-law fits to the Sombrero metal-poor and metal-rich subpopulations yields exponents of $\alpha=-2.1\pm0.1$ and $\alpha=-1.9\pm0.1$, respectively.  Fainter than the bend, power-law fits to the subpopulations and the entire sample are consistent with a flat distribution.

Observational evidence suggests the GCLF of ellipticals galaxies and Local Group spirals (M31 and the Milky Way) share a universal TOM \citep{h01, ric03}.  A composite GCLF from Virgo Cluster dwarf ellipticals (dEs) appears to have a similar TOM to massive Virgo galaxies \citep{sea05B}.  Table~\ref{tabdistmod} shows distance estimates measured to the Sombrero galaxy from a variety of sources.  If the midpoint between the surface brightness fluctuation (SBF) and the planetary nebulae luminosity function distance is adopted (i.e. m$-$M$=29.77\pm0.03$), then the ACS absolute TOM is M$^{TOM}_{V} = -7.60\pm0.06$.  

After applying an correction of $-0.16$ mag to account for a re-calibration of the \citet{tea01} SBF distances (as discussed in Jensen et al. 2003) the \citet{ric03} ``universal'' GCLF TOM value is M$^{TOM}_{V} = -7.35$, with a dispersion of 0.25 mag.  This is almost three tenths of a magnitude fainter than the Sombrero TOM.  A slight discrepancy such as this may originate in GCLF TOM estimation uncertainties like contamination, completeness limits, photometric calibration, and varying GCLF fitting techniques \citep{ric03}.  The distance moduli are uncertain as well, therefore the Sombrero TOM is likely consistent with the universal value.  If instead the Sombrero TOM is assumed to correspond exactly to the universal TOM, a distance modulus of m$-$M = 29.52 (8.0~Mpc) is derived.

\begin{figure}
\center
\includegraphics[scale=0.4,angle=0]{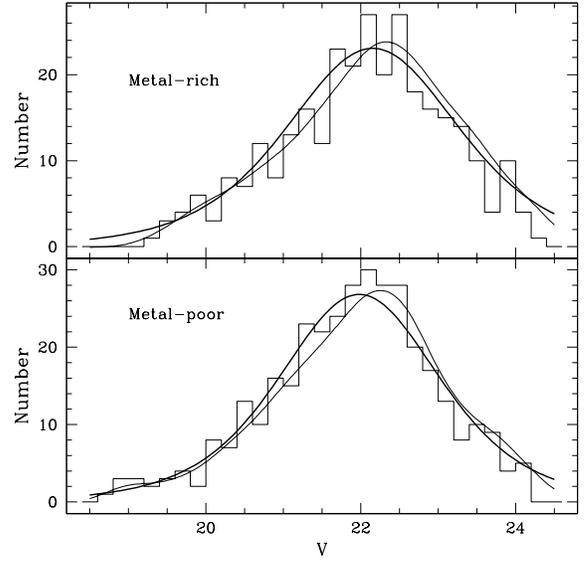}
\caption{GC subpopulation luminosity functions.  Plots are the same as in Figure~\ref{figlfv}, with metal-poor (m$^{TOM}_{blue} = 22.11\pm0.08$) and metal-rich (m$^{TOM}_{red} = 22.20\pm0.09$) subpopulations in the bottom and top panels, respectively.\label{figsubpoplf}}
\endcenter
\end{figure}

GCLFs of the subpopulations are shown in Figure~\ref{figsubpoplf}.  In contrast to what has been found observationally in several galaxies \citep[e.g.][]{lea01} and demonstrated theoretically \citep{acz95}, the blue TOM in the Sombrero is not brighter than the red TOM:  $\Delta V^{TOM}_{blue-red}=-0.09\pm0.12$ mag (m$^{TOM}_{blue} = 22.11\pm0.08$, $\sigma_{blue}=1.08\pm0.08$, m$^{TOM}_{red} = 22.20\pm0.09$ and $\sigma_{red}=1.16\pm0.08$; TOMs are from the peak of a density estimate, while the TOM error and dispersion values are from the Secker algorithm).  The predicted difference for metallicity peaks of [Fe/H]$ = -1.4$ and [Fe/H]$ = -0.4$, and assuming the two subpopulations have the same old ages, GC mass functions, and stellar initial mass function is $\Delta V_{blue-red}^{TOM} = -0.30$ \citep{acz95}.  The TOM differences from other bands are also marginally smaller than expected:  $\Delta B^{TOM}_{blue-red}=-0.27\pm0.12$ versus the predicted $\Delta B^{TOM}_{blue-red}=-0.52$ and $R^{TOM}_{blue-red}=0.04\pm0.12$ versus the predicted $\Delta R^{TOM}_{blue-red}=-0.25$.

The difference between the observed and predicted values may indicate that the simplifying assumptions made about the GCLFs in \citet{acz95} do not accurately represent the Sombrero subpopulations.  For example, according to the \citet{bc03} SSP models if the metal-rich GCs are younger than the metal-poor GCs by 1 Gyr then from age effects alone a metal-rich GC is expected to be brighter by 0.04 V-band magnitudes compared to a metal-poor GC of the same mass.

Within the mosaic region of complete radial coverage a TOM difference between the inner ($0\arcmin \leq$ R$_{gc} < 1.5\arcmin$) and outer ($1.5\arcmin \leq $ R$_{gc} < 3.0\arcmin$) GCLFs is not statistically significant according to the \citet{sec92} algorithm:  $\Delta V^{TOM}_{in-out}=0.11\pm0.13$.  When the subpopulations are each divided into the same radial bins, a statistically significant TOM difference is not observed among either subpopulation:  $\Delta V^{TOM}_{in-out}=-0.09\pm0.25$ and $\Delta V^{TOM}_{in-out}=0.23\pm0.19$ for the blue and red GCs, respectively.  For each of these radial comparisons, the corresponding differences in GCLF dispersion values are found to be insignificant.  A null result for such analysis suggests that various radial-dependant dynamical evolution mechanisms \citep[e.g. ][]{fz01,vea03} do not shape the Sombrero GCLF in an observable manner.

\subsection{Blue Tilt}\label{sectiontilt}

A visual inspection of the color-magnitude diagram in Figure~\ref{figgoodcmd} suggests a correlation exists between color and magnitude for metal poor GCs, while no such relation is apparent for the metal-rich GCs.  Quantifying this ``blue tilt'' requires careful consideration of the biases due to parameter limits, photometric errors, and contaminants.  For instance, the introduction of a subpopulation boundary at a single color (i.e. blue GCs strictly have B$-$R~$\leq 1.3$) will bias an examination of the individual subpopulation color-magnitude distributions.  For this reason, the Nmix routine (introduced in $\S$\ref{sectioncmd}) is used to quantify the subpopulation color peaks at five magnitude intervals.  Two lines were then fitted to the blue and red Nmix peaks with a least-squares minimization routine that was weighted by the Nmix peak standard deviation in the mean.  Weighting the color peaks by this value effectively propagates photometric error information, since less weight will be placed on the higher color dispersion characteristic of fainter GCs.

The magnitude intervals were chosen to cover the entire luminosity range while still maintaining a numerical significance at each interval.  Since the faint end of the magnitude distribution is more affected by photometric uncertainty, the faintest magnitude interval is excluded from the final linear fit but is shown in Figure~\ref{figgoodcmd}.  Indeed, whether the faintest bins follows the trends established by the brighter points is indeterminate.  Table~\ref{tabtilt} summarizes the slopes from linear fits to all band and color combinations.

The formal significance of the linear slopes from the B$-$R Nmix peak fits are consistent with the visual impression that a more significant slope is present for the blue GCs, $\delta$(B$-$R)/$\delta$R $=-0.035\pm0.005$, compared to the red GCs:  $\delta$(B$-$R)/$\delta$R $=-0.015\pm0.007$ (see the fits in Figure~\ref{figgoodcmd}).  From Table~\ref{tabtilt} it is evident that for a given color, the slope (and the formal significance) changes depending on which band is used on the abscissa.  When the midpoint between the red subpopulation $\delta$(B$-$R)/$\delta$B and $\delta$(B$-$R)/$\delta$R slopes is considered, $-0.012\pm0.007$, it provides no strong evidence for an analogous red GC trend.

Two recent ACS studies of GC systems discovered similar blue tilts in the metal-poor subpopulation.  \citet{sea05B} showed that the massive Virgo cluster galaxy, M87 (and perhaps NGC 4649, but not NGC 4472), has a metal-poor color-magnitude trend in the GCs brighter than GCLF TOM.  The Sombrero trend extends to GC at least V $\sim 22.7$ (M$_V \sim -7.1$), consistent with the Sombrero GCLF TOM and the magnitude range found in \citet{sea05B}.  \citet{hea05} found similar trends in eight brightest cluster/group galaxies.  However, they suggested that the blue tilt is found only among the brightest of the blue GCs and gradually begins to disappear at a transitional magnitude approximately 1.5 mag brighter than the GCLF TOM.  Neither \citet{sea05B} nor \citet{hea05} found any evidence for an color-magnitude trend among the metal-rich GCs.  The discovery of a blue tilt in a field spiral like Sombrero shows this phenomena is not restricted to the GC systems of massive ellipticals in high-density galaxy environments.  Blue tilts have since been discovered in other spiral galaxies (Forde~et~al.~in~prep.).

\citet{hea05} converted their color-magnitude trend to a metal-poor GC metallicity-mass gradient:  $Z\propto M^{0.55}$.  \citet{sea05B} found similar values of $Z\propto M^{0.48}$ in M87 and $Z\propto M^{0.42}$ in NGC~4649.  The corresponding gradient from the Sombrero data, $Z\propto M^{0.27}$, is from the midpoint of the metallicity-mass proportionalities from the $\delta$B$-$R/$\delta$B and $\delta$B$-$R/$\delta$R color-magnitude slopes.  Systematics from band and color-to-metallicity transformation differences cannot be ruled out as the cause for the smaller Sombrero gradient.  However, a similar color-to-metallicity transformation technique to \citet{hea05} was used here, supporting a real difference in the trends.

Two interpretations of the GC metallicity-mass trend were proposed in the earlier studies:  ``pre-enrichment'', where larger proto-GC clouds contain higher metal abundances and ``self-enrichment'' where GCs were able to retain and then incorporate some metals created within the GC.  \citet{hea05} supported a pre-enrichment origin after noting that some numerical simulations predict that only the most massive dE galaxies ($> 10^9$M$_\sun$) can retain metals, while anything less massive will tend to eject metal-rich gases \citep[e.g.,][]{mlf99,fea04}.  It was suggested by \citet{hea05} that dEs in this mass regime are capable of building GCs with masses well above $10^6$M$_\sun$, a value which corresponds to slightly more than half a magnitude brighter than their transitional magnitude of M$_I = -10.0$ (they assume M/L $=3$).  The Sombrero and \citet{sea05B} results support a fainter transitional magnitude ($\sim 1.5$ mag fainter) indicating the trend is apparent among objects less massive than the mass limit for GC pre-enrichment quoted in \citet{hea05}.

In terms of \textit{absolute} metallicity, metal-rich GCs have on average 10 times higher metallicity than metal-poor GCs, thus any given metallicity-magnitude gradient found among metal-poor GCs would correspond to a 10 times smaller slope for the metal-rich GCs, which may have disappeared in the noise.  If the metal-rich GCs actually do not follow an analogous trend, different formation environments (e.g. Strader et al.~2006 and Harris et al.~2006) or subpopulation characteristics (e.g. differing radial distributions) are both conceivable ideas that would require careful theoretical modelling to test.  Indeed, without predictions from modelling of individual GCs formation and evolution, attempts to explain physically the processes responsible for this new discovery will largely remain speculative.

\subsection{Radial Distribution} \label{raddist}

\begin{figure}
\center
\includegraphics[angle=-90,scale=0.33]{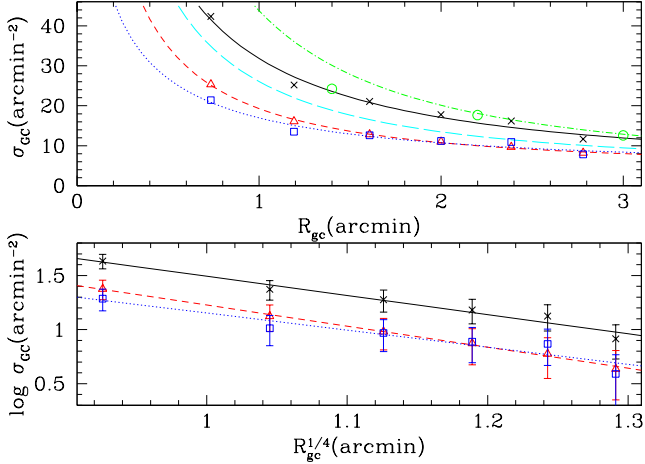}
\caption{The GC surface densities of the entire population (crosses) and the two subpopulations (square and triangle for blue and red GCs, respectively).  The solid, dotted, and short-dashed lines are de~Vaucouleurs profile fits to the entire, blue and red surface densities, respectively.  The lower panel is a log representation while the top panel shows a linear view.  In the upper panel the dash-dotted line and circle symbols are from ground-based analysis of the Sombrero GC surface density \citep{rz04} and the long-dashed line is from HST WFPC2 \citep{lfb01} analysis.  Error bars are from Poisson statistics.  For clarity, the online version of this Figure is presented in color.\label{figsurfdensdeva}}
\endcenter
\end{figure}

\begin{figure}
\center
\includegraphics[angle=-90,scale=0.33]{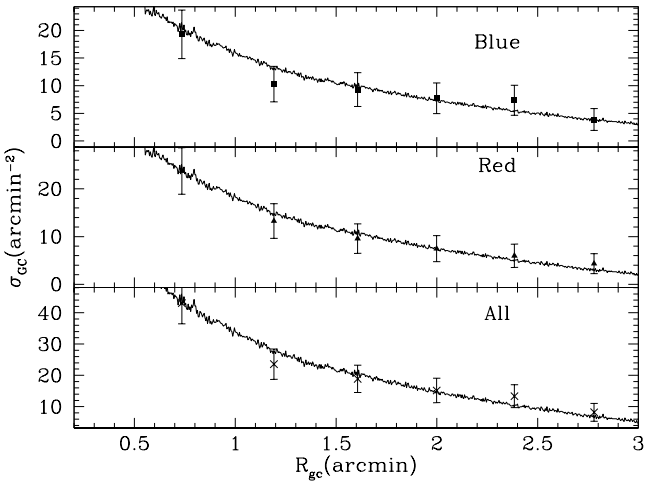}
\caption{Surface densities of the blue subpopulation (top), red (middle) and the entire GC system (bottom).  These are compared with the arbitrarily normalized R-band galaxy luminosity profile.  The GC surface densities all resemble the galaxy profile.\label{figsurfdens}}
\endcenter
\end{figure}

The Sombrero metal-rich GCs are more centrally-concentrated than their metal-poor counterparts, as shown in Figure~\ref{figsurfdensdeva}.  From Figure~\ref{figsomb} it is visually apparent the metal-rich GCs are associated with the Sombrero bulge rather than the disk, confirming the work of \citet{mea02} and suggestions by \citet{fbl01}.  The surface densities of the entire GC system and the two subpopulations show slopes which are similar to the galaxy light profile (Figure~\ref{figsurfdens}) with neither subpopulation displaying a particularly better match.

Two functions traditionally used to model GC surface densities ($\sigma_{GC}$) are a power-law, log($\sigma_{GC}$)$ = \alpha_{pow}\cdot $log(R$_{gc}$)$+\beta_{pow}$, and de~Vaucouleurs law, log($\sigma_{GC}$)$ = \alpha_{deV}\cdot R_{gc}^{1/4} + \beta_{deV}$.  A shallower power-law exponent, $\alpha_{pow} = -1.14\pm0.13$, is found compared to the surface density exponent derived recently from ground-based, wide-field imaging:  $\alpha_{pow} = -1.85\pm0.07$ \citep{rz04}.  The de~Vaucouleurs fit also yields a shallower curve of $\alpha_{deV} = -1.77\pm0.17$ compared to ground-based and HST WFPC2 \citep{lfb01} analysis, which found slopes of $\alpha_{deV} = -2.11\pm0.08$ and $\alpha_{deV} = -2.00\pm0.18$, respectively.  The masked Sombrero disk region was excluded from the surface density calculation.

The current ACS and past WFPC2 de~Vaucouleurs surface density fits to the entire GC population are compared in Figure~\ref{figsurfdensdeva}.  Given that the WFPC2 fit was derived from only three pointings (thus inherently uncertain due to both the small areal coverage and GC sample size), it is likely consistent with the ACS results within the uncertainty due to sampling statistics.  While the fit from the \citet{rz04} is clearly offset from the ACS curve, the surface density points from this past study are consistent with the ACS fit.  This supports the idea that a shallower surface density is present in the central galaxy regions compared to regions outside of the ACS field of view.  \citet{bh92} found that Sombrero profile could be best represented by a composite of two power laws transitioning at $2.5\arcmin$ with a shallower profile in the inner radial region.  This is also consistent with observations of other GC systems that exhibit flattened cores in their GC radial distributions \citep[e.g. ][]{dea03}.

\begin{figure}
\center
\includegraphics[angle=-90,scale=0.33]{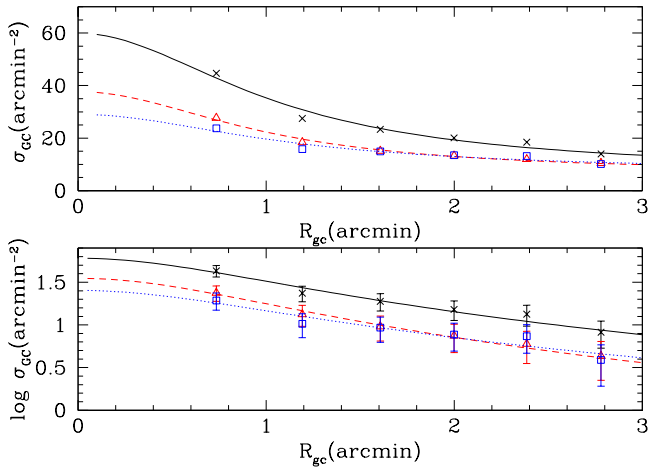}
\caption{The GC surface densities.  Same as Figure~\ref{figsurfdensdeva} except a ``cored'' profile (presented in $\S\ref{raddist}$) is fit to the surface densities.  This profile provides a more accurate population estimate because, unlike the de Vaucouleurs profile, it does not continue to increase significantly as R$_{gc}\rightarrow0$.\label{figsurfdenscore}} 
\endcenter
\end{figure}

Adopting an outer radius of $19\arcmin$ (where $\sigma_{GC} \sim 0$ arcmin$^{-2}$; Rhode \& Zepf 2004) and integrating the de Vaucouleurs surface density profile, a total GC population of $1870\pm850$, a blue total of $1110\pm850$ and a red total of $820\pm200$ were estimated (population errors are $1\sigma$ values from Monte-Carlo simulations of the integration of the surface density profile).  The total population derived here is consistent with the 1900 GCs estimated by \citet{rz04}.  However, the de Vaucouleurs profile continues to increase significantly as R$_{gc} \rightarrow 0$ and may result in an overestimation of the total.  Indeed, the de Vaucouleurs profile of the entire GC population overestimates the number of GCs found in the innermost radial intervals (N$^{predicted}_{GC}\approx130$ in $0.3\arcmin\leq$R$_{gc}<1.0\arcmin$) compared to the number of observed GCs (N$^{observed}_{GC}\approx90$) by $\approx30\%$.  Fitting a ``cored'' profile of the form $\sigma_{GC} = \alpha\cdot(1+($r$_{gc}/1\arcmin)^2)^\beta$ (from the empirical surface density fit in Dirsch et al.~2003; see Figure~\ref{figsurfdenscore}) to the surface densities of entire population, and the blue and red subpopulations separately, yields the reduced estimates of $1540\pm640$, $940\pm620$, and $670\pm270$, respectively (errors computed analytically).

\citet{fbl01} showed that the number of red GCs normalized to the bulge luminosity, the bulge specific frequency (bulge S$_N$), has a similar value in the Milky Way, M31 and the Sombrero, perhaps indicating a common bulge formation history.  To compute the bulge S$_N$ a fraction of the total galaxy luminosity must be attributed to the bulge component.  The Sombrero reddening-corrected total B$_T$ magnitude is $8.39\pm0.06$ and the corrected color is B$-$V$ = 0.87\pm0.01$ (RC3), which corresponds to $M_V = -22.25\pm0.06$ for a distance modulus of $29.77\pm0.03$.  Adopting a bulge-to-total luminosity ratio of 0.73 \citep{bba98}, resulted in a bulge apparent magnitude of $M_V = -21.91\pm0.06$.

A bulge S$_N = 1.2\pm0.8$ was calculated (errors from the GCs totals) from the extrapolated ``cored'' red surface density fit (N$_{red}=670\pm270$).  This value is larger than the bulge S$_N$ estimated for M31, the Milky Way \citep{fbl01}, and five spiral galaxy GC systems analyzed in a WFPC2 compilation by \citet{gea03}.  In these previous studies a higher degree of uniformity was found if instead a bulge S$_N$ was calculated from the number of red GCs within two effective radii (R$_{1/2}$) of the galaxy bulge.  Adopting a Sombrero bulge effective radius of $0.89\arcmin$ \citep{bba98}, $160\pm30$ red GCs within 2R$_{1/2}$ are estimated, which corresponds to a bulge S$_N = 0.3\pm0.05$.  This version of the Sombrero bulge S$_N$ is consistent with M31 \citep[bulge S$_N = 0.4\pm0.2$;][]{fbl01} and is larger than all of the spirals studied in \citet{gea03}.

The M31 bulge S$_N$ is likely more uncertain than the formal errors quoted above and an overestimate since the number of metal-rich GCs is derived mostly from photographic plate analysis \citep{fea93}, which is poorly suited to address the significant amount of contamination expected because of the face-on orientation of the M31 disk.  Thus the Sombrero bulge S$_N$ value is likely larger than all the spiral GC systems on which this analysis has been performed.  
If spirals generally have a constant bulge S$_{N}$, it would support the idea that most metal-rich GCs formed in conjunction with their host bulge and the Sombrero analysis would seem to suggest that something characteristic of more massive galaxies (e.g. mass-to-light ratios, formation timescales) leads to enhanced metal-rich GC formation and/or decreased destruction rates.

\begin{figure}
\center
\includegraphics[angle=0,scale=0.4]{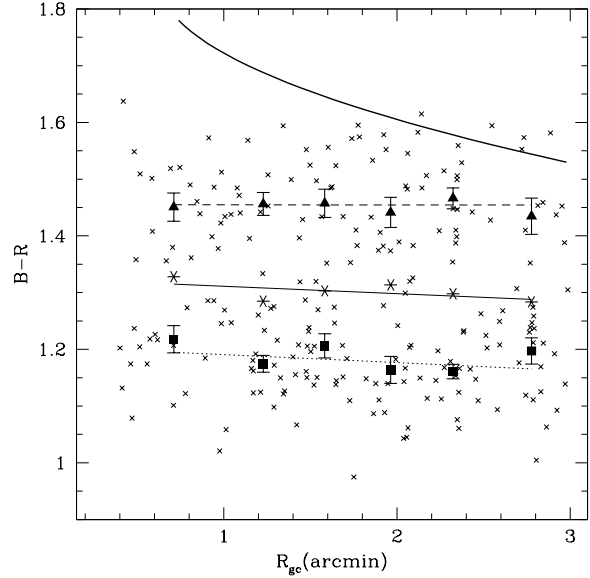}
\caption{GC color versus projected galactocentric radius.  Crosses are those GCs with V $< 22$ magnitudes.  No statistically significant gradient is found in either the red (dashed line and triangles) or blue subpopulations (dotted line and squares).  Error bars are the standard deviation in the means of the peak color estimates.  The mean B$-$R values for the total GC population (solid line and stars) illustrate the changing ratio of blue to red GCs with radius.  The thick solid line shows a fit to the ACS mosaic galaxy color profile.\label{figradicolor}}
\endcenter
\end{figure}

When GC Nmix peak colors derived from GCs brighter than V $=22.0$ are plotted against R$_{gc}$ (Figure~\ref{figradicolor}), no significant trend is observed in the red ($-0.000\pm0.015$ mag arcmin$^{-1}$) and blue subpopulation ($-0.014\pm0.012$ mag arcmin$^{-1}$).  Figure~\ref{figradicolor} also shows that the galaxy color profile extracted from the ACS mosaic is redder than the metal-rich subpopulation but begins to show GC-like colors at $\sim2\arcmin$.

\subsection{Sizes}\label{sizesection}

Figure~\ref{figrawsize} hints at a size-color trend for the GCs. Histograms of the half-light radii of the blue and red GC subpopulations are presented in Figure~15. The blue and red subpopulations have median R$_{hl}$ of 2.12~pc and 1.87~pc, respectively; thus the reds are systematically $\approx13\%$ smaller than the blue GCs (mean sizes 2.40~pc and 2.10~pc; reds $\approx14\%$ smaller). Using the same size determination algorithm (ISHAPE) and a Sombrero WFPC2 F547M-band image centered on the galaxy center, \citet{lea01} found mean half-light radii of 2.16~pc and 1.67~pc (updated with the current distance modulus) for the blue and red subpopulations, respectively, corresponding to a larger size difference of about 30\%.  If the ACS size measurements are restricted to the GCs found in the same region as the central WFPC2 image (see Figure~\ref{figsomb}), a blue-red mean size difference of 30\% is found, with mean R$_{hl}$ of 2.25~pc and 1.73~pc for the blue and red GCs, respectively.  This size difference is consistent with the WFPC2 data, suggesting that the smaller average size difference for the full ACS mosaic is a real effect, due to the larger area covered by the ACS data.  

\begin{figure}
\center
\includegraphics[angle=0,scale=0.4]{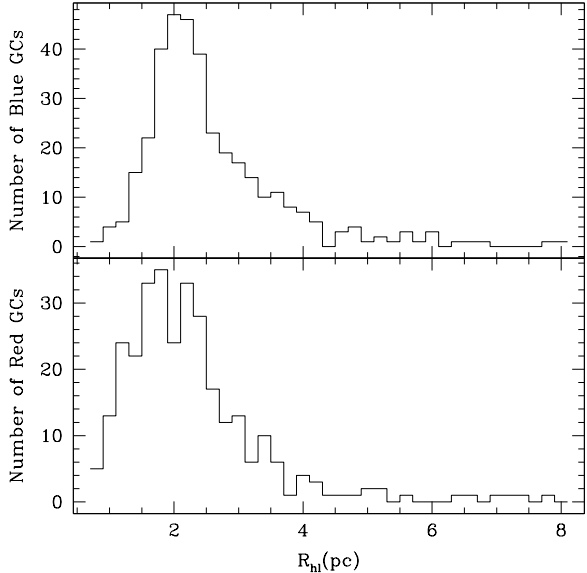}
\caption{GC half-light radius histograms for blue (top panel) and red (bottom panel) GC subpopulations.  The red GCs appear 13\% smaller on average than the blue GCs with median R$_{hl}$ values of 2.12 and 1.87~pc for the blue and red subpopulations, respectively.\label{figsizhist}}
\endcenter
\end{figure}

Two main interpretations of the size difference between blue and red GCs have been proposed.  \citet{lb03} suggested typical observations of the central component of a GC system could lead to a subpopulation size difference from the projection of the different GC subpopulation radial distributions (with the red GCs being more centrally-concentrated) and an underlying GC size dependence on the three-dimensional galactocentric radius (R$_{3D}$) of all GCs.

Alternatively, in \citet{j04} it was shown that a correlation of GC half-light radii and metallicity can arise from mass segregation combined with a metallicity dependent main-sequence lifetime.  In the same study, it was shown that a form of this model fit observations of M87 GCs.  It is important to note that the size-color trends predicted by the \citet{j04} models are strongly dependant on the relative ages of the two subpopulations.  To reproduce observations \citet{j04} assumed the subpopulations are coeval.  While GC system formation scenarios generally place metal-rich GC formation later than the metal-poor GCs and the ages of Milky Way GC may show an age-metallicity relationship \citep[e.g. ][]{sw02}, a precise spectroscopic constraint on extragalactic GC ages is not currently achievable \citep{sea05A}. 

\citet{jea05} addressed both of these theories in their analysis of a large sample of GC systems from early-type galaxies in the ACS Virgo Cluster Survey \citep[VCS; ][]{cea04}.  They concluded the intrinsic metallicity difference between blue and red GCs provides a better explanation for the observed GC subpopulation size differences than projection effects largely because the size-R$_{gc}$ trends they observed were suggested to be too shallow to cause the observed mean size difference.  However, the size-R$_{gc}$ trends they examined were restricted to the metal-poor subpopulation, which is more affected by projection effects compared to metal-rich GCs as demonstrated below.

Irrespective of the previous results, \citet{j04} noted the only way to unequivocally distinguish between the two theories is to examine GCs at large projected radii, since the GCs furthest from the projected galaxy center will be more likely to share a similar R$_{3D}$ value, thus enabling a comparison between the subpopulations without the influence of an underlying size-R$_{3D}$ trend.

\begin{figure}
\center
\includegraphics[angle=-90,scale=0.33]{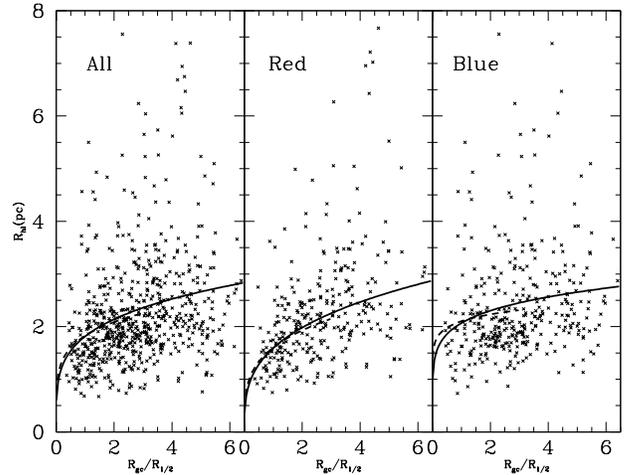}
\caption{GC sizes versus galactocentric radius normalized by the Sombrero bulge effective radius.  Left, middle, and right sections contain the entire population, red subpopulation, and blue subpopulation, respectively.  Solid lines are power-law fits to the entire radial coverage while the dashed lines are fits only to the inner GCs (R$_{gc}/$R$_{1/2} < 3$).  The difference between the fit to the inner and entire blue GC sample suggests projection effects are more apparent in the central regions for this subpopulation.\label{figsizerad}}
\endcenter
\end{figure}

The ACS mosaic coverage extends to $\approx6$ times the Sombrero bulge effective radius \citep[R$_{1/2}=0.89\arcmin$;][]{bba98}.  This is roughly twice the radial limit reached by \citet{jea05} in their most massive galaxies, whose populous GC systems constituted the bulk of their size sample.  Figure~\ref{figsizerad} shows the clear GC size-(R$_{gc}/$R$_{3D}$) trends in the Sombrero subpopulations and the entire population.  When a power-law relation of the form log(R$_{hl}$)$= a + b\cdot$log(R$_{gc}$/R$_{1/2}$) \citep[as in ][]{jea05} is fit to the data, the following slopes are derived:  $b_{blue} = 0.16\pm0.04$, $b_{red} = 0.32\pm0.05$, and $b_{all} = 0.24\pm0.03$.  For comparison, \citet{jea05} found for the metal-poor GCs an average weighed slope of $b^{virgo}_{blue} = 0.07\pm0.01$ from their sample of Virgo GC systems.  Fitting curves to the Sombrero size-R$_{gc}$ information restricted to the Virgo radial limit (R$_{gc}$/R$_{1/2}<3$) yields shallower exponents compared to the unrestricted ACS values and a blue subpopulation value which is consistent with the \citet{jea05} value:  $b_{blue} = 0.09\pm0.07$, $b_{red} = 0.27\pm0.07$, and $b_{all} = 0.19\pm0.05$.  

\begin{figure}
\center
\includegraphics[angle=-90,scale=0.33]{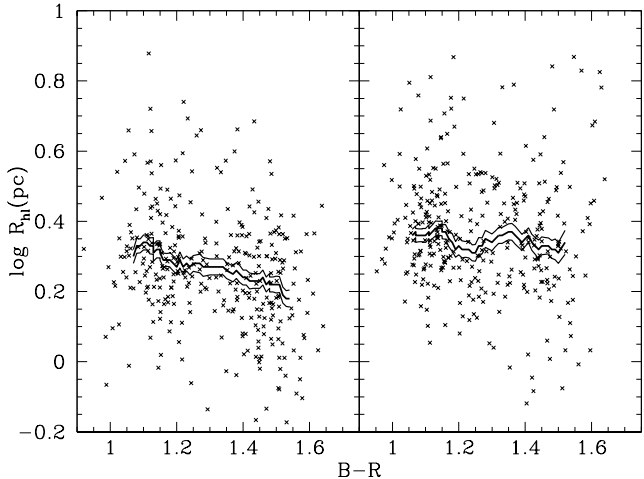}
\caption{Size-color plots from two equally-sized galactocentric radius intervals divided at R$_{gc} = 2.4\arcmin$.  The solid lines represent moving biweight location estimators from 50 data values with 1$\sigma$ boundaries shown.  In contrast to the inner radial interval (left panel), the biweight values from the outer interval (right panel) are consistent with no size-color trend.  The percentage difference between the subpopulation sizes is $17\pm5\%$ and $5\pm6\%$ for the inner and outer intervals, respectively.  These observations are most likely a result of projection effects and the different radial distribution of the two subpopulations.\label{figcolsizerad}}
\endcenter
\end{figure}

Figure~\ref{figcolsizerad} effectively contrasts both the GC subpopulation size theories and clearly illustrates that there is little subpopulation size difference at larger R$_{gc}$.  This result was alluded to at the beginning of this section when the ACS sizes were compared to the results from the Sombrero WFPC2 image, which was centered on the galaxy (see Figure~\ref{figsomb}).  The nearly flat blue GC size-R$_{gc}$ gradient in the inner region ($<3$ R$_{gc}$/R$_{1/2}$; Figure~\ref{figsizerad}) could be caused by the introduction of scatter from the wider range of blue GC R$_{3D}$ values (i.e., projection effects).  These observations suggest projection effects are responsible for the appearance of a Sombrero subpopulation size difference.

To conclude, the observed Sombrero subpopulation size difference is consistent with the data in \citet{jea05}, although the extended radial coverage leads to a different interpretation.  Projection effects are most likely responsible for the observed average size difference between blue and red GCs in the Sombrero and that, in general, projection effects should always be accounted for when comparing GC sizes from different GC systems.  The Sombrero size analysis reaffirms that GC size information at large R$_{gc}$ is essential for disentangling the two interpretations of GC subpopulation size difference.  Also, it is important to examine the metal-rich subpopulation for size-R$_{gc}$ trends since this subpopulation is less susceptible to projection effects from its centrally-concentrated distribution.  

\subsection{Size-Magnitude Trend}\label{sectionsizemag}

\begin{figure}
\center
\includegraphics[scale=0.33,angle=-90]{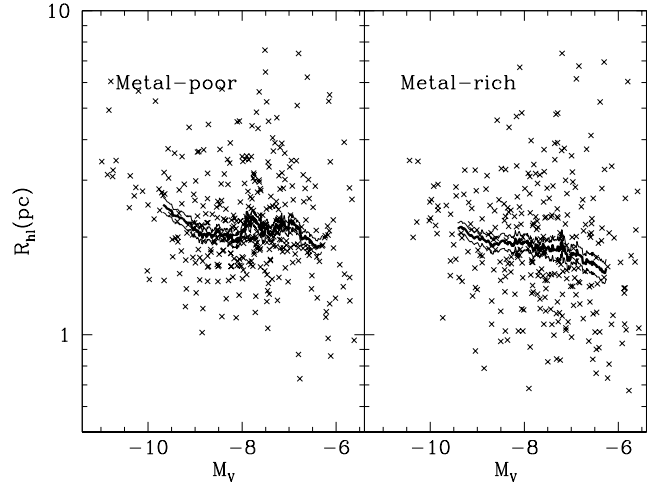}
\caption{Globular cluster subpopulation half-light radii versus magnitude.  The line is a moving biweight location estimator (width of 50 data points) with 1$\sigma$ boundaries.  The metal-poor GCs show no clear size-magnitude trend among the fainter GCs until M$_V \leq -9.0$ mag where a strong upturn is observed.  The metal-rich GCs are consistent with this upturn and also show increasing sizes with luminosity along the entire magnitude range.\label{figmagsiz}}
\endcenter
\end{figure}

Figure~\ref{figmagsiz} shows the size-magnitude diagrams of both GC subpopulations.  Solid lines show moving biweight location estimator values from the ROSTAT software routine \citep{bea90} with a window width of 50 data points.  Among the metal-poor GCs, no clear size-magnitude trend is detected from the faintest GC magnitudes to M$_V =-9.0$ where a strong size-magnitude ``upturn'' is observed such that the brighter GCs show increasingly larger sizes on average.  The metal-rich GCs exhibit a shallower size-magnitude trend over the entire magnitude range where the average sizes again become larger as the GC luminosity increases.

In $\S\ref{sizeselect}$ it was shown that the $1\sigma$ uncertainties from a linear fit to the size-size plots derived from each of the three image bands corresponds to an accuracy level of $\sim0.2$~pc (Figure~\ref{figbandsizes}), which is approximately half the dynamical range of the size-magnitude trends.  When the same analysis is performed on GCs of three different magnitude intervals, it is estimated that the sizes are accurate at the 0.15, 0.17, and 0.27~pc levels for the brightest (M$_V\leq-8.3$), intermediate ($-8.3<$ M$_V\leq-7.4$) and faintest (M$_V>-7.4$) thirds of luminosity distribution, respectively.

\begin{figure}
\center
\includegraphics[scale=0.4,angle=0]{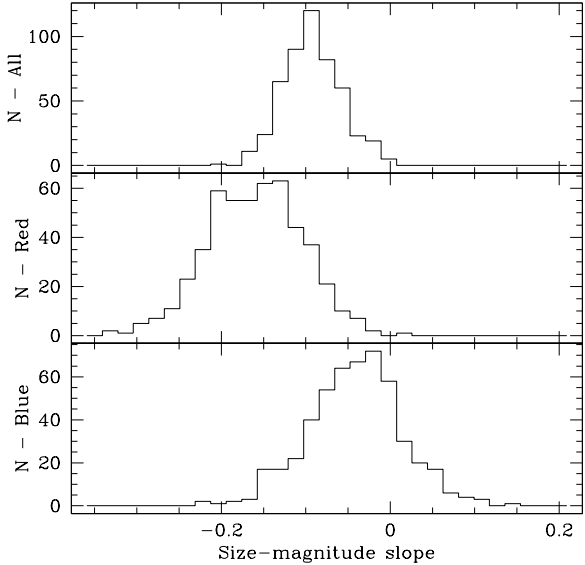}
\caption{Histogram of slopes from a bootstrap statistical analysis of the size-magnitude trends for the blue, red, and entire GC sample in the bottom, middle and top panels, respectively.  The high-luminosity size-magnitude upturn was intentionally avoided in this analysis, as described in the text.  This analysis suggests a significant size-magnitude trend exists among the metal-rich GCs and marginally so among the metal-poor GCs.\label{figsizmagboot}}
\endcenter
\end{figure}

To determine the whether the shallow trend is significant, the two subpopulations and the entire sample were bootstrapped and divided into magnitude intervals of approximately 50 data points.  The biweight of each interval was calculated and a least-squares linear fit performed on the resultant biweight values for all but the brightest magnitude interval to prevent the size-magnitude upturn from dominating the fit.  Figure~\ref{figsizmagboot} contains histograms from the output of this procedure and suggests a significant trend is present among the metal-rich GCs and marginally so in the metal-poor subpopulation.

\begin{figure}
\center
\includegraphics[scale=0.4,angle=0]{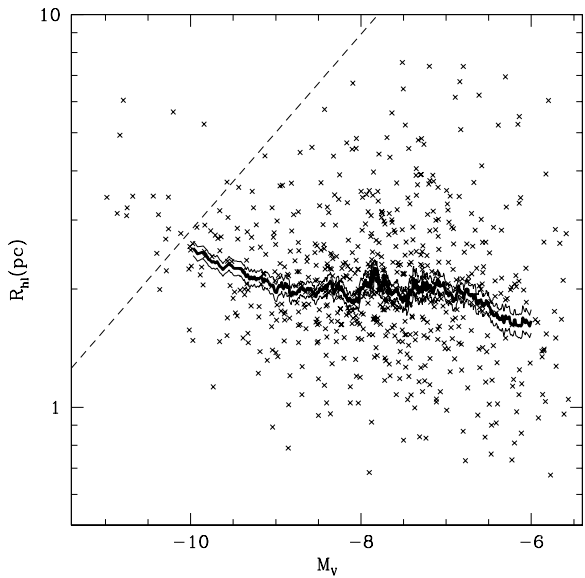}
\caption{Globular cluster half-light radii versus magnitude.  The line is a moving biweight location estimator (width of 50 data points) with 1$\sigma$ boundaries indicated by the lines.  The dashed line represents the division between Milky Way GCs and the suspected Local Group accreted dE,N nuclei candidates \citep{mvdb05}.\label{figmagsiz2}}
\endcenter
\end{figure}

\subsubsection{Size-Magnitude Upturn and More Massive Stellar Systems}

To date, no positive size-luminosity trend has been reported in any extragalactic GC systems \citep[e.g.][]{kw01,lea01}, although there exists no comparably large dataset of GC sizes from such high signal-to-noise images as is presented here.  There are, however, a growing number of studies in which very luminous and extended GCs are being found.  For example in NGC 5128 fourteen massive (M$_V > -9.6$; M $> 10^6$ M$_\sun$) GCs were shown to have unusually large sizes \citep{mh04,harea02}.  \citet{mh04} hypothesized these to be accreted nucleated dwarf elliptical (dE,N) nuclei that were stripped of their halos.  Using ACS data \citet{hea05} found 20 to 30 bright and extended objects in NGC 1407.  Also certain Galactic GCs brighter than M$_V = -9.4$ ($\omega$ Cen, M54, and NGC 2419) are all larger than the median Milky Way GC size and are suggested to be accreted dE,N nuclei \citep[see refs. in ][]{mvdb05}.  Figure~\ref{figmagsiz2} shows that the boundary between those suspected Local Group accreted dE,N nuclei and typical Milky Way GCs \citep{mvdb05} approximately corresponds to where the size-magnitude upturn appears in the Sombrero.

These observations support the idea that a group of particularly massive and extended GCs may exist in other galaxies, indicating the Sombrero high-luminosity GC size-magnitude upturn may not be unique.  It is also apparent that utilizing an accreted dE,N nuclei scenario to explain the presence of massive and extended GC-like objects is common practise.  This idea is supported by a resemblance between GCs (metal-poor GCs in particular) and nuclei still embedded within their dE,N galaxy.  Similarities between massive, metal-poor GCs and dE nuclei include:  broadband colors \citep{lea04,depea05,hea05}, the peak in the dE nuclei luminosity function \citep[M$_V\sim-9.5$; from table~2 of ][]{lea04} corresponds to the most luminous GCs (and where the Sombrero GC size-magnitude upturn appears), and dE nuclei and metal-poor GCs show a nearly identical color-magnitude trend \citep{hea05}.

\citet{cea06} studied the nuclei of 51 early-type galaxies from the ACS VCS and showed that the sizes of the nuclei increase with their luminosity.  If the nuclei size-magnitude trend is partially preserved during the accretion of a dE,N onto the Sombrero, a size-magnitude trend might begin to appear as the number of accreted nuclei becomes comparable to the high-luminosity indigenous metal-poor GCs.

Another possible explanation for the size-magnitude upturn is that it reflects an intrinsic property of GC formation.  This idea is motivated by an analogous size-mass upturn found in the young counterparts to GCs, young massive clusters \citep{kpjb05}.  Supporting this is a striking absence in Figure~\ref{figmagsiz2} of Sombrero GCs with M$_V<-10$ that are smaller than the median size of the sample (R$_{hl}\approx2$~pc).  This is unexpected given the intrinsic scatter in sizes is quite large (Sombrero size standard deviation is $\sim1$~pc) and if the upturn was caused entirely by the presence of accreted dE nuclei, high-luminosity indigenous GCs with small sizes should still be observed.  Thus an absence of small bright objects is consistent with an intrinsic origin for the size-magnitude upturn.

Similarly determining if a deficiency of bright GCs with small sizes exists in other systems (and therefore a size-magnitude upturn) is a simple exercise to test whether the size-magnitude upturn is a common feature.  An examination of figure~6 in \citet{mea06} containing the large GC size-magnitude dataset from the ACS Fornax Cluster Survey (Jordan et al., in prep.) suggests a such deficiency exists in other GC systems.

\citet{mea06} proposed that all compact stellar systems (in particular Ultra Compact Dwarfs, UCDs) with M$_V<-11.0$ show increasing sizes with increasing luminosity, while less luminous stellar systems have magnitude-independent sizes and therefore are likely to be typical GCs.  The figure that partially motivated this idea is the same which here is suggested to demonstrate a deficiency of luminous objects with small sizes.  Thus the size-magnitude upturn shown in figure~6 of \citet{mea06} may actually begin at M$_V\lesssim-10.0$ where the deficiency noted here becomes apparent.  A fainter absolute magnitude marking the emergence of the Sombrero upturn, M$_V\lesssim-9.0$, may indicate the upturn appears in even less massive compact stellar systems than demonstrated in \citet{mea06}.  If the Sombrero upturn is indeed an intrinsic feature of its massive GCs, it supports a continuous trend of magnitude-dependent sizes from the brightest GCs to more massive stellar systems such as dE nuclei and UCDs.

\subsubsection{Metal-Rich GC Size-Magnitude Trend}

Validating the shallower metal-rich size-magnitude trend is difficult because completeness effects (where faint, extended GCs are missed) conceivably could have caused the trend.  A quantitative investigation into these effects and a careful consideration of the numerous size-related trends \citep[as demonstrated in ][]{jea05} is beyond the scope of the present work.  It is also noted that a combination of the extended radial distribution of the metal-poor subpopulation and an underlying size-R$_{gc}$ trend (i.e., projection effects, see $\S\ref{sizesection}$) plausibly could introduce enough scatter to hide an analogous size-magnitude trend in the metal-poor subpopulation.

Even so, similar size-magnitude trends have been found among young star clusters of the merger galaxy NGC 3256 \citep{zea99} and 18 nearby spiral galaxies \citep{l04}.  The former study found that the clusters follow a size-luminosity proportionality of R$_{hl} \propto L^{\sim0.07}$, compatible with the results of the later spiral study.  The corresponding Sombrero metal-rich GC fit, R$_{hl} \propto L^{0.07\pm0.03}$, is identical to the past results.  A fit to biweight location estimates of five Sombrero metal-rich GC magnitude intervals confirm this analysis:  R$_{hl} \propto L^{0.09\pm0.01}$.  While it is curious that the results are consistent across the different studies, it is still indeterminate whether the consistency is a reflection of a real trend or the result of a similar type of completeness effect.

\section{Summary and Conclusions}

A six-pointing mosaic from the HST Advanced Camera for Surveys was used to study the globular cluster system of the Sombrero galaxy.  An unprecedentedly deep ($\sim95\%$ of GCLF and $\sim2$ mags fainter than the turnover magnitude), virtually contamination-free sample, allowed an accurate GC luminosity function (TOM) turnover magnitude to be determined:  m$_V^{TOM} = 22.17\pm0.06$ or M$_V^{TOM} = -7.60\pm0.06$, assuming m$-$M$=29.77\pm0.03$.  The GCLF was found to be well fit by a t$_5$ to 2 magnitudes below the TOM (Figure~\ref{figlfv}).  The dispersion of the t$_5$ fit is $\sigma^{ACS}_{t5} = 1.08\pm0.06$.  While the TOM is slightly fainter than a universal TOM, M$^{TOM}_{V} = -7.35$ \citep{ric03,jea03}, it is likely consistent within the typical observational errors.  No variation with R$_{gc}$ is found in the Sombrero GCLF TOM, nor is a significant difference between the subpopulation TOMs detected.

The GC color distribution is clearly bi-modal (Figures~\ref{figrawcmd} and \ref{figgoodcmd}) with blue and red peaks at B$-$R $= 1.15$ and $1.44$ or [Fe/H] = $-1.38$ and $-0.49$, respectively (see Table~\ref{tabcolor}).  Subpopulation peak colors are consistent with the expected values for a galaxy with a luminosity like the Sombrero, according to the relation between the GC subpopulation color peaks and host galaxy luminosity \citep{sbf04}.  It was found that the color dispersion in the metal-poor subpopulation is smaller compared to the metal-rich subpopulation.  This is in agreement with the results of Harris et al. (2006), Peng et al. (2006), and Strader et al. (2006).

An estimated 53\% of this GC sample belongs to the blue subpopulation.  This proportion contrasts to the results from ground-based observations where a larger proportion of 59\% was found from data that covers the entire spatial distribution of the GC system \citep{rz04}.  Thus the difference is likely caused by the decreasing ratio of red to blue GCs with galactocentric radius (Figure~\ref{figradicolor}).  No color-R$_{gc}$ trend is observed within either subpopulation.  The galaxy color profile tends to be redder than metal-rich GCs and only begins to overlap this subpopulation at R$_{gc}>2\arcmin$.

The best-fitting de~Vaucouleurs radial profile of the Sombrero GC system from \citet{rz04} is steeper than the one found here.  However, their surface density values from within the ACS mosaic region are consistent with the ACS surface densities (Figure~\ref{figsurfdensdeva}).  This observation is consistent with there being a core of constant number density in the inner GC system, confirming observations made in an earlier Sombrero GC system study \citep{bh92}.

The number of red GCs normalized to the bulge luminosity, the bulge specific frequency, is found to be larger than all spiral galaxies for which this value has been calculated.  This suggests the rate of metal-rich GC formation was enhanced and/or GC destruction was less effective in the Sombrero galaxy relative to less massive spirals.

A color-magnitude trend, a so-called ``blue tilt'', was discovered in the metal-poor GC subpopulation that is likely a metallicity-mass trend (Figure~\ref{figgoodcmd}).  The metal-rich subpopulation shows no strong evidence for an analogous trend.  The Sombrero provides the first example of this trend in a spiral galaxy and a galaxy found in a low-density galaxy environment.  The blue tilt extends from the brightest GCs to at least the GCLF TOM, in agreement with the findings of \citet{sea05B}.  A shallower trend, in terms of metallicity-mass proportionality, is found here compared to examples in massive ellipticals found in high-density galaxy environments \citep{sea05B,hea05}.  This difference and the absence of a blue tilt in NGC~4472 likely indicates the gradient of the color-magnitude trend depends on the specific properties of the host galaxy.

It is shown that GC sizes correlate with R$_{gc}$ (Figure~\ref{figsizerad}).  From the extended radial coverage provided by the Sombrero ACS mosaic, the metal-rich GCs are found to have a steeper size-R$_{gc}$ gradient compared to the metal-poor subpopulation, thus causing the mean subpopulation sizes to become identical at larger R$_{gc}$ (Figure~\ref{figcolsizerad}).  These observations are consistent with projection effects being largely responsible for the observed GC subpopulation size difference \citep{lb03}, as opposed to an intrinsic feature of their respective metallicities \citep{j04}.

A size-magnitude upturn was discovered among the brightest GCs (M$_V<-9.0$), where the sizes increase as a function of increasing luminosity (Figure~\ref{figmagsiz}).  The accretion of dwarf elliptical nuclei into the Sombrero GC system provides a plausible scenario for this trend.  However, an analogous trend among young massive clusters \citet{kpjb05} and a noticeable deficiency of luminous GCs (M$_V<-10.0$) with smaller than average sizes (Figure~\ref{figmagsiz2}; also see figure~6~in Mieske et al. 2006), supports an intrinsic origin for the size-magnitude upturn.  An intrinsic size-magnitude upturn suggests the most massive GCs are at the low-mass end of a continuous trend that includes more massive systems such as dwarf nuclei and UCDs.

A shallower positive size-magnitude trend among metal-rich GCs is found at all magnitudes.  The luminosity-size proportionality, R$_{hl} \propto L^{0.07\pm0.03}$, is consistent with the same fits to young star clusters of the merger galaxy NGC 3256 \citep{zea99} and 18 nearby spiral galaxies \citep{l04}.  It is unclear whether the consistency is a reflection of a real trend or the result of a completeness effect.

\acknowledgments
\section{Acknowledgments}

We thank the anonymous referee for a thorough reading and comments that greatly improved the quality of the text.  We appreciate helpful discussions with R. Proctor.  This work was supported by NSF grant AST-0507729 and JS acknowledges a Graduate Research Fellowship.  Support for this work was provided by NASA through grant number G0-09766 from the Space Telescope Science Institute, which is operated by AURA, Inc., under NASA contract NAS 5-26555.  DF thanks the ARC for financial support.

\clearpage

\begin{deluxetable}{lccc}
\tablewidth{0pt}
\tablecolumns{4}
\tablecaption{Aperture and Reddening Corrections\label{tbl-corr}}
\tablehead{
\colhead{Filter} & \colhead{F435W (B)} & \colhead{F555W (V)} &
\colhead{F625W (R)}}
\startdata
AC(5-10 pixels) & 0.187 & 0.191 & 0.198 \\
AC(10-infinity) & 0.107 & 0.092 & 0.088 \\
Total AC & 0.294 & 0.283 & 0.286 \\
Mean Extinction Correction (A$_X$)  & & & \\
from the six pointings & 0.210 & 0.163 & 0.136 \\
\enddata
\end{deluxetable}

\begin{deluxetable}{lccccccccc}
\tablewidth{0pt}
\tablecolumns{10}
\tablecaption{Color/Metalicity Distributions and Subpopulation Proportions\label{tabcolor}}
\tablehead{
\colhead{} & \colhead{Blue} & \colhead{Red} & \colhead{Blue [Fe/H]} & \colhead{Red [Fe/H]} & \colhead{$\sigma_{blue}$} & \colhead{$\sigma_{red}$} & \colhead{$\sigma^{[Fe/H]}_{blue}$} & \colhead{$\sigma^{[Fe/H]}_{red}$} & \colhead{Blue \%} }
\startdata
 B$-$R & 1.15 & 1.44 & $-1.32$ & $-0.46$ & 0.08 & 0.10 & 0.23 & 0.29 & 53\% \\
 B$-$V & 0.67 & 0.90 & $-1.41$ & $-0.50$ & 0.05 & 0.07 & 0.24 & 0.31 & 52\% \\
 V$-$R & 0.46 & 0.55 & $-1.24$ & $-0.52$ & 0.02 & 0.03 & 0.22 & 0.32 & 52\% \\
\enddata
\end{deluxetable}

\begin{deluxetable}{lccl}
\tablewidth{0pt}
\tablecaption{Sombrero Distance Moduli Estimates\label{tabdistmod}}
\tablehead{ \colhead{Technique (source)} & \colhead{m$-$M} & \colhead{Mpc} & \colhead{Notes}}
\startdata
Surface Brightness  & $29.79\pm0.18$ & $9.1\pm0.8$ & Re-calibrated by  \\
Fluctuations (SBF) & & & $-0.16$ mag. \\
\citep{tea01} & & & \citep{jea03}\\
\hline
 Faber-Jackson & 30.03 & 10.1 & B$_T = 9.41\pm0.02$ \\
relation & & & $\sigma = 237\pm6$\\
(HyperLeda) & & & m$-$M$=$B$_T+6.2$log$\sigma+5.9$\\
\hline
Virgo infall  & 30.85 & 14.8 & Close proximity makes\\
velocity  & & & velocity uncertain\\
(HyperLeda) & & & H$_0=72$km/s/Mpc\\
\hline
Planetary Nebulae  & $29.74^{+0.04}_{-0.06}$ & $8.9\pm0.6$ & \\
LF (PNLF) & & & \\
\citep{fea96} & & & \\
\hline\hline \\

Adopted distance & $29.77\pm0.03$ & $9.0\pm0.1$ & Weighted average of\\
to Sombrero & & &  SBF \& PNLF \\
\enddata
\end{deluxetable}

\begin{deluxetable}{ccccccc}
\tablewidth{0pt}
\tablecaption{GC Color-Magnitude and Metallicity-Mass Gradients\label{tabtilt}}
\tablehead{  \colhead{color/band}  & \colhead{Blue Color} & \colhead{Red Color} & \colhead{ $Z\propto M^{\alpha}$}}
\startdata
(B-R)/R & $-0.038\pm0.005$ & $-0.015\pm0.007$ & $\alpha=0.29$ \\
(B-R)/B & $-0.032\pm0.005$ & $-0.007\pm0.007$ & $\alpha=0.24$ \\ 
(B-V)/V & $-0.032\pm0.004$ & $-0.015\pm0.005$ & $\alpha=0.35$ \\
(B-V)/B & $-0.027\pm0.004$ & $-0.007\pm0.005$ & $\alpha=0.30$ \\
(V-R)/R & $-0.009\pm0.002$ & $-0.004\pm0.002$ & $\alpha=0.21$ \\
(V-R)/V & $-0.010\pm0.002$ & $-0.004\pm0.002$ & $\alpha=0.24$ \\
\enddata
\end{deluxetable}

\end{document}